\documentclass[%
aps,prd,
showpacs,twocolumn,notitlepage,
amssymb,amsmath,amsfonts,mathrsfs,
nofootinbib,superscriptaddress,
floats,floatfix
]{revtex4-2}

\usepackage{amsmath}
\usepackage{graphicx}
\usepackage{epstopdf}
\usepackage{enumerate}
\usepackage{soul}

\usepackage{savesym}
\savesymbol{tablenum}
\usepackage{siunitx}
\restoresymbol{SIX}{tablenum}

\usepackage{hyperref}

\newcommand{\mnras}{MNRAS}
\newcommand{\apjs}{ApJS}
\newcommand{\apjl}{ApJL}
\newcommand{\aap}{A\&A}

\newcommand{\beq}{\begin{equation}}
\newcommand{\eeq}{\end{equation}}
\newcommand{\beqn}{\begin{eqnarray}}
\newcommand{\eeqn}{\end{eqnarray}}
\newcommand{\pa}{\partial}

\def\2pi{2\pi}

\begin{document}


\title{Supernova-like explosion of massive rotating stars from disks surrounding a black hole}

\author{Sho Fujibayashi}
\affiliation{Max-Planck-Institut f\"ur Gravitationsphysik (Albert-Einstein-Institut), Am M\"uhlenberg 1, D-14476 Potsdam-Golm, Germany}

\author{Alan Tsz-Lok Lam}
\affiliation{Max-Planck-Institut f\"ur Gravitationsphysik (Albert-Einstein-Institut), Am M\"uhlenberg 1, D-14476 Potsdam-Golm, Germany}

\author{Masaru Shibata}
\affiliation{Max-Planck-Institut f\"ur Gravitationsphysik (Albert-Einstein-Institut), Am M\"uhlenberg 1, D-14476 Potsdam-Golm, Germany}
\affiliation{Center for Gravitational Physics and Quantum Information, Yukawa Institute for Theoretical Physics, Kyoto University, Kyoto, 606-8502, Japan}

\author{Yuichiro Sekiguchi}
\affiliation{Center for Gravitational Physics and Quantum Information, Yukawa Institute for Theoretical Physics, Kyoto University, Kyoto, 606-8502, Japan}
\affiliation{Department of Physics, Toho University, Funabashi, Chiba 274-8510, Japan}


\date{\today}


\begin{abstract}
We perform a new general-relativistic viscous-radiation hydrodynamics simulation for supernova-like explosion associated with stellar core collapse of rotating massive stars to a system of a black hole and a massive torus paying particular attention to large-mass progenitor stars with the zero-age main-sequence mass of $M_\mathrm{ZAMS}=20$, 35, and $45M_\odot$ of Ref.~\cite{Aguilera-Dena2020oct}. Assuming that a black hole is formed in a short timescale after the onset of the stellar collapse, the new simulations are started from initial data of a spinning black hole and infalling matter that self-consistently satisfy the constraint equations of general relativity. It is found that with a reasonable size of the viscous parameter, the supernova-like explosion is driven by the viscous heating effect in the torus around the black hole irrespective of the progenitor mass. The typical explosion energy and ejecta mass for the large-mass cases ($M_\mathrm{ZAMS}=35$ and $45M_\odot$) are $\sim 10^{52}$\,erg and $\sim 5M_\odot$, respectively, with  $^{56}$Ni mass larger than $0.15M_\odot$. These are consistent with the observational data of stripped-envelope and high-energy supernovae such as broad-lined type Ic supernovae. This indicates that rotating stellar collapses of massive stars to a black hole surrounded by a massive torus can be a central engine for high-energy supernovae. By artificially varying the angular velocity of the initial data, we explore the dependence of the explosion energy and ejecta mass on the initial angular momentum and find that the large explosion energy $\sim 10^{52}$\,erg and large $^{56}$Ni mass $\geq 0.15M_\odot$ are possible only when a large-mass compact torus with mass $\agt 1M_\odot$ is formed. 
\end{abstract} 


\maketitle

\section{Introduction}
\label{sec:intro}

Gravitational-wave observations by advanced LIGO and advanced Virgo have shown that stellar-mass black holes with a wide mass range between $\sim 3M_\odot$ and $\sim 100M_\odot$ are commonplace in the universe~\cite{Abbott2021apr,Abbott2021nov}. It is natural to consider that a majority of these black holes are formed from core collapse of massive stars. In particular for large black-hole mass, $M_\mathrm{BH}\agt 20M_\odot$, the black holes are likely to be formed shortly after the stellar core collapse with a short proto-neutron star stage or directly during the stellar core collapse. However, it is still not very clear how these black holes are formed. One way to understand the formation process of the black holes is to detect electromagnetic signals emitted during the formation and subsequent evolution processes such as gamma-ray bursts~\cite{Woosley1993,Piran2004}. However, the observational information of the stellar center is limited because the formed black hole is hidden by the dense matter surrounding it. Therefore, to understand the formation and evolution processes of the black holes during the stellar core collapse, theoretical studies play a crucial role.  

A numerical-relativity simulation incorporating the relevant physics such as neutrino transfer, equation of state for high-density matter, and angular-momentum transport is the chosen way to theoretically understand the formation and evolution processes of stellar-mass black holes. In our previous paper~\cite{Fujibayashi2022dec}, we performed numerical-relativity simulations with approximate neutrino transfer and shear viscous hydrodynamics employing relatively low-mass (9 and $20M_\odot$), compact, rotating progenitor stars derived by stellar evolution calculations of Ref.~\cite{Aguilera-Dena2020oct}. We showed that these stars collapse to a black hole shortly after the formation of a proto-neutron star and subsequently the black holes grow due to the mass accretion from the infalling envelope. In the long-term (several seconds) evolution, an accretion disk is developed due to the centrifugal force of late-time infalling matter. The disk subsequently becomes a geometrically thick torus by the effects of viscous heating, viscous angular momentum transport, and shock heating. During an early stage in which the neutrino cooling efficiency and the ram pressure by the infalling matter are high, the outflow of the matter from the torus is prohibited. However, in a later stage, the neutrino cooling efficiency and the ram pressure become low enough to induce the mass outflow from the system, leading to a supernova-like explosion for the entire progenitor star (see also Ref.~\cite{Just2022aug} for a related work). 

The previous work~\cite{Fujibayashi2022dec} also showed that the explosion energy may be larger than that of the typical supernovae if the progenitor stars are rapidly rotating and a high mass-infall rate onto the torus is achieved. In such a case, a compact and massive ($\agt 1M_\odot$) disk/torus can be formed around a black hole and the viscous and shock heating on the disk/torus can provide a large amount of the thermal energy, which can be the source for an energetic explosion. The viscous heating rate in a disk is written approximately as $\dot E_\nu \sim \nu M_\mathrm{torus} \Omega^2$ with the torus mass $M_\mathrm{torus}$, angular velocity $\Omega$, and shear viscous coefficient $\nu$. In the alpha viscous prescription~\cite{Shakura1973a}, $\nu$ is written as
\beqn
\nu=\alpha_\nu c_\mathrm{s} H,
\eeqn
where $\alpha_\nu$ is the so-called alpha parameter, $c_\mathrm{s}$ is the sound velocity, and $H$ is the scale height of the torus approximately written as $H=c_\mathrm{s}/\Omega$. Then, the viscous heating rate is 
\beqn
\dot E_\nu &\sim& 4 \times 10^{52}\,{\rm erg/s}\,
\left(\frac{\alpha_\nu}{0.03}\right)
\left(\frac{M_\mathrm{torus}}{M_\odot}\right)\nonumber \\
&\times&\left(\frac{c_\mathrm{s}}{10^9\,{\rm cm/s}}\right)^2
\left(\frac{M_\mathrm{BH}}{10M_\odot}\right)^{-1/2}
\left(\frac{R}{10 M_\mathrm{BH}}\right)^{-3/2},\label{eq:edot_vis}
\eeqn
where we used $\Omega\approx\sqrt{M_\mathrm{BH}/R^3}$ with $M_\mathrm{BH}$ and $R$ being the black hole mass and cylindrical radius of the torus. Here, the viscosity is supposed to be induced effectively by magnetohydrodynamics turbulence; see e.g., Refs~\cite{Balbus:1998ja,Hawley:2013lga,Suzuki:2013rka,Shi:2015mvh,kiuchi2018a,Held:2022gds,Hayashi:2021oxy}, which shows $\alpha_\nu=O(10^{-2})$. 
In the presence of matter infall onto the disk/torus, strong shear layers are also formed at the shock surfaces outside the disk/torus, and hence, the viscous heating can be even more enhanced.

The timescale of the viscous heating in the disk/torus is written as 
\beqn
t_\nu&:=&{R^2 \over \alpha_\nu c_\mathrm{s} H} \nonumber \\
&\approx &4.7 \,{\rm s}\,
\left({\alpha_\nu \over 0.03}\right)^{-1}
\left({c_\mathrm{s} \over 10^9\,{\rm cm/s}}\right)^{-2}\nonumber \\
&\times & \left({M_\mathrm{BH} \over 10 M_\odot}\right)^{1/2}
\left({R \over 10 M_\mathrm{BH}}\right)^{1/2},
\eeqn
and thus, the total dissipated energy is approximately
\beqn
\dot E_\nu t_\nu &\sim & {M_\mathrm{torus}M_\mathrm{BH} \over R} \nonumber \\
&\approx& 1.8 \times 10^{53}\,{\rm erg}
\left({M_\mathrm{torus} \over M_\odot}\right)
\left({10 M_\mathrm{BH} \over R}\right).
\eeqn
Hence, if a fraction of the energy released by the viscous heating contributes to the outflow of the matter, it is possible to achieve a supernova-like explosion with a very large explosion energy of order $10^{52}$\,erg in the presence of a compact and large-mass torus of $M_\mathrm{torus}\sim 0.1$--$1M_\odot$.

In this paper, we continue our exploration of this problem for more massive progenitor stars with zero-age main-sequence mass $M_\mathrm{ZAMS}=35$ and $45M_\odot$ as well as $M_\mathrm{ZAMS}=20M_\odot$. Following our previous work, we employ the stellar evolution models by Aguilera-Dena et al.~\cite{Aguilera-Dena2020oct}. Since these stars have compact and very massive cores at the onset of the collapse, we may expect formation of a black hole shortly after the core bounce~\cite{Oconnor2011apr} (but see Ref.~\cite{Burrows2019} for a counter example). In this work, therefore, we assume the black-hole formation after the core bounce without an explosion in the proto-neutron star stage. Under this assumption, we prepare an initial condition composed of a spinning black hole and infalling matter that self-consistently satisfy constraint equations of general relativity. The initial condition is prepared for a stage with no accretion disk/torus formation.  With such initial data, we perform a neutrino-radiation viscous hydrodynamics simulation in full general relativity paying particular attention to the disk/torus formation and evolution, and subsequent development of the matter outflow, which leads to a supernova-like explosion.

This paper is organized as follows: In Sec.~\ref{secII}, we summarize the progenitor models which we employ and then describe how to set up the initial condition composed of a spinning black hole and infalling matter. 
Section~\ref{sec:results} presents the results of numerical-relativity simulations focusing on the mechanism of the explosion, the explosion energy, the ejecta property, and predicted light curves of the supernova-like explosion. Section~\ref{secV} is devoted to a summary. In Appendix~\ref{A1}, we describe a formulation for the initial-value problem of general relativity that we employ in this paper. In Appendixes~\ref{A2} and \ref{A3}, supplemental numerical results are presented. Throughout this paper we basically use the geometrical units of $c=1=G$ where $c$ and $G$ are the speed of light and gravitational constant, respectively, but when it is necessary to clarify the units, we recover $G$ and $c$. $k_\mathrm{B}$ denotes Boltzmann's constant. 

\section{Models and initial conditions}\label{secII}

We employ massive and very compact progenitor stars among the stellar evolution models of Ref.~\cite{Aguilera-Dena2020oct}. Specifically, we select the stars with the mass of the zero-age main-sequence state, $M_\mathrm{ZAMS}=20$, 35, and $45M_\odot$. For these stars, we may suppose that a black hole would be formed in a short timescale after the core bounce because the compactness parameter of Ref.~\cite{Oconnor2011apr} is very large.
\footnote{Even for extremely compact progenitor stars, a supernova explosion may occur and a black hole may not be formed via neutrino heating~\cite{Burrows2019} and/or via magnetohydrodynamics effects~\cite{Burrows2007,Obergaulinger2021,Obergaulinger2022may}, although our previous simulations for the $20M_\odot$ progenitor model indicate that the assumption of the black-hole formation may be valid for the progenitor models of Ref.~\cite{Aguilera-Dena2020oct}.}

Assuming the conservation of the specific angular momentum during the formation and subsequent growth of a black hole, it is possible to approximately determine the mass and angular momentum of the formed black hole for a given profile of the specific angular momentum as a function of the enclosed mass $j(m)$~\cite{Shibata:2002br,Shibata:2003iw}, if the region with the enclosed mass $m$ collapses to the black hole without forming a disk. In the following, we assume that the angular velocity profile $\Omega$ is a function of spherical radius only, as is done in the stellar evolution calculation~\cite{Aguilera-Dena2020oct}, and thus, the specific angular momentum $j$ represents the angular average as
\beq
    j = \frac{1}{4\pi r^2}\int_0^{2\pi} \int_0^\pi \Omega(r) r^4 \sin^3 \theta d\theta d\varphi = \frac{2}{3} r^2 \Omega(r).
\eeq
Since $j$ is a function of $r$, $m$ is as well. 

Then, we choose the mass of the black hole, $M_\mathrm{BH,0}$, which is much larger than the maximum mass of neutron stars of $\alt 3M_\odot$. The resulting angular momentum, $J_\mathrm{BH,0}$, of the black hole is written as
\beq
J_\mathrm{BH,0}=\int_0^{M_\mathrm{BH,0}} j(m') dm'.
\eeq
We note that for the choice of $M_\mathrm{BH,0}$, $j(m)$ with any value of $m\leq M_\mathrm{BH,0}$ has to be smaller than the specific angular momentum of the innermost stable circular orbit $j_\mathrm{ISCO}$~\cite{Bardeen:1972fi} of the black hole of mass $m$ and angular momentum 
\beq
J(m)=\int_0^m j(m') dm'.
\eeq
Since the angular momentum of the black hole is determined by specifying the enclosed mass, $j_\mathrm{ISCO}$ is a function of the enclosed mass $m$ in this context.

\begin{figure}
\includegraphics[width=0.48\textwidth]{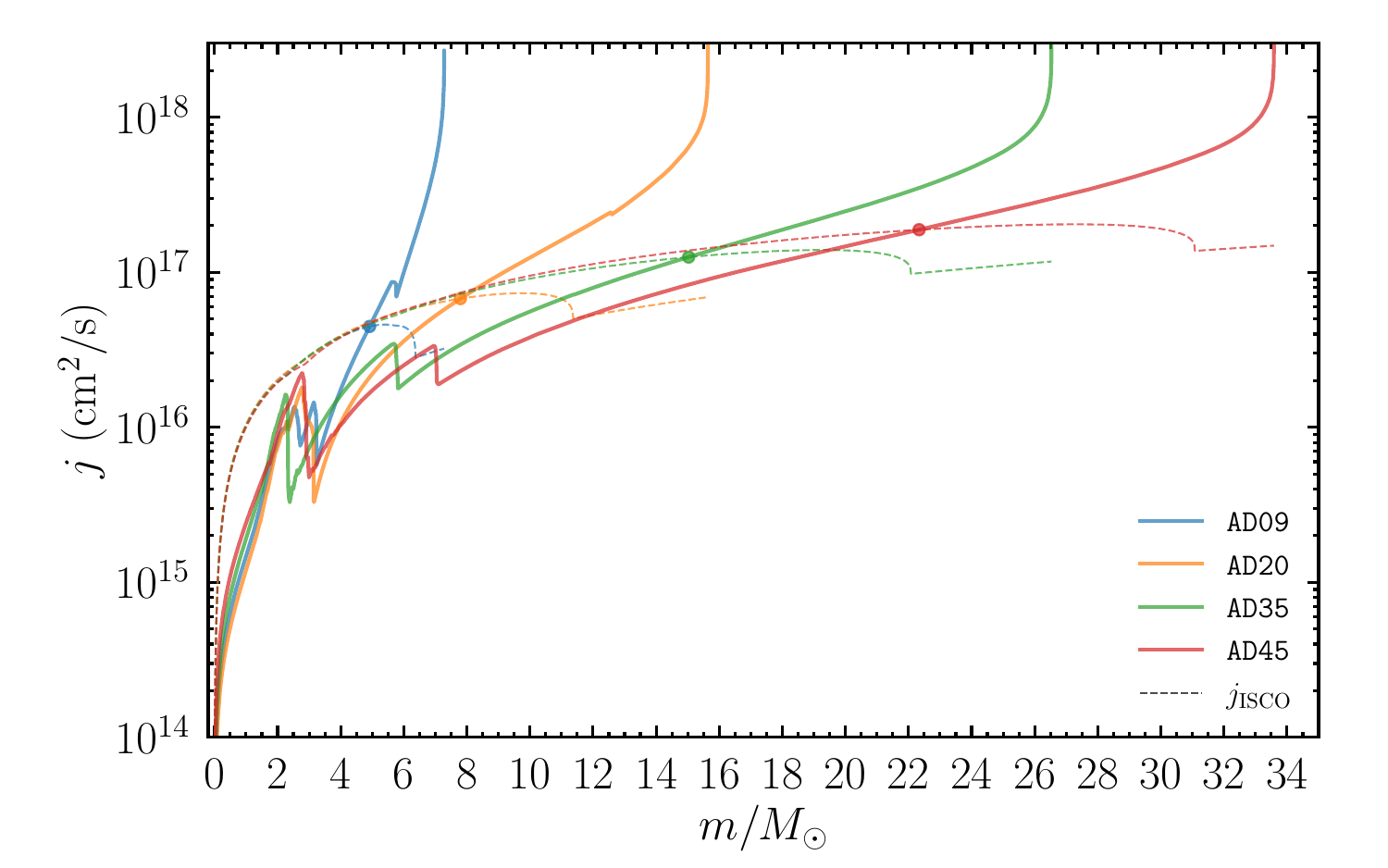}
\caption{Specific angular momentum, $j$, as a function of the enclosed mass, $m$, for the models of $M_\mathrm{ZAMS}=9$, 20, 35, and $45M_\odot$ in  Ref.~\cite{Aguilera-Dena2020oct} (solid curves). We also plot $j_\mathrm{ISCO}$ for a given black hole of mass $m$ and corresponding angular momentum $J(m)$ by the dotted curves. The filled circles denote the points at which $j=j_\mathrm{ISCO}$ is satisfied for each stellar model. 
}
\label{fig1}
\end{figure}

Figure~\ref{fig1} shows $j$ as a function of $m$ for $M_\mathrm{ZAMS}=9$, 20, 35, and $45M_\odot$ of Ref.~\cite{Aguilera-Dena2020oct} (solid curves). We also plot $j_\mathrm{ISCO}$ by the dotted curves. The filled circles denote the points at which $j=j_\mathrm{ISCO}$ is satisfied (we refer to the corresponding mass as  $M_\mathrm{ISCO}$). This figure shows that for any model, $j(m) < j_\mathrm{ISCO}$ is satisfied for $m < M_\mathrm{ISCO}$ and indicates that for the progenitor models with $M_\mathrm{ZAMS}=20$, 35, and $45M_\odot$, a black hole is likely to grow to $M_\mathrm{BH}=M_\mathrm{ISCO} \approx 8$, 15, and $22M_\odot$ prior to the disk formation. In the presence of the viscous angular-momentum transport, the disk formation is delayed and black holes with larger mass can be formed before the disk formation.

\begin{table*}[t]
    \centering
    \caption{Model description. Model name, mass of the zero-age main-sequence  stars, $M_\mathrm{ZAMS}$, employed angular velocity profile, initial rest mass (including the fraction which is transformed to the black hole), initial mass and dimensionless spin of the black hole, the ratio of the matter angular momentum $J_\mathrm{mat}$ to the black-hole angular momentum $J_\mathrm{BH,0}=M_\mathrm{BH,0}^2\chi_0$, alpha parameter for viscosity, and grid spacing for the central region, $\Delta x_0$, respectively. The last two columns present the mass and dimensionless spin of the black hole at the termination of the simulations. Note that for model \texttt{AD20-7.8}, we stopped the simulation on the way of further significant black-hole growth (see Fig.~\ref{fig:BH}). The results for model \texttt{AD20x1} are taken from Ref.~\cite{Fujibayashi2022dec}.
    }
    \begin{tabular}{lccccccccccccccc}
    \hline\hline
        Model & $M_\mathrm{ZAMS}$  & $\Omega$ profile & ~~$M_{*,0}$~~ & ~~$M_\mathrm{BH,0}$~~ & ~~$\chi_0$~~ & $J_\mathrm{mat}/J_\mathrm{BH,0}$& ~~$\alpha_\nu$~~  & $\Delta x_0$\,(m)& ~$M_\mathrm{BH,f}$~ & ~$\chi_\mathrm{BH,f}$~ \\
        \hline
        \texttt{AD20-7.8} & $20M_\odot$ & original &$15.1M_\odot$& ~$7.8M_\odot$~ & 0.60 & 9.93 & 0.03 & 250 & $(10.4M_\odot)$ & (0.74)\\
        \texttt{AD20-9} & $20M_\odot$ & original &$15.1M_\odot$& $9.0M_\odot$ & 0.72 & 5.60 &0.03 & 216 & $10.8M_\odot$ & 0.79 \\
        \texttt{AD20-10} & $20M_\odot$ & original &$15.0M_\odot$& $10.0M_\odot$ & 0.83 & 3.86 &0.03 & 240 & $10.9M_\odot$ & 0.84 \\
        \texttt{AD35-15} & $35M_\odot$ & original &$25.5M_\odot$& $15.0M_\odot$ & 0.66 & 4.32 &0.03 & 360 & $20.2M_\odot$ & 0.81\\
        \texttt{AD35-15-hi} & $35M_\odot$ & original &$25.4M_\odot$& $15.0M_\odot$ & 0.66 & 4.53 & 0.03 & 300 & $19.6M_\odot$ & 0.81\\
        \texttt{AD35-15-mv} & $35M_\odot$ & original &$25.5M_\odot$& $15.0M_\odot$ & 0.66 & 4.33 & 0.06 & 360 & $19.6M_\odot$ &0.79\\
        \texttt{AD35-15-hv} & $35M_\odot$ & original &$25.5M_\odot$& $15.0M_\odot$ & 0.66 & 4.32 & 0.10 & 360 & $18.9M_\odot$ &0.78\\
        \texttt{AD35x0.5-21.5} & $35M_\odot$ & original$\times0.5$~  &$25.5M_\odot$& $21.5M_\odot$ & 0.48 & 0.84 & 0.03 & 516 & $25.1M_\odot$ & 0.60\\
        \texttt{AD35x0.6-21.5} & $35M_\odot$ & original$\times0.6$~  &$25.5M_\odot$& $21.5M_\odot$ & 0.58 & 0.84 & 0.03 & 516 & $24.5M_\odot$ & 0.66\\
        \texttt{AD35x0.8-18} & $35M_\odot$ & original$\times0.8$~  &$25.4M_\odot$& $18.0M_\odot$ & 0.63 & 2.13 & 0.03 & 432 & $22.2M_\odot$ & 0.75\\
        \texttt{AD35x1.2-12.5} & $35M_\odot$ & original$\times1.2$~  &$25.5M_\odot$& $12.5M_\odot$ & 0.69 & 8.18 & 0.03 & 300 & $18.2M_\odot$ & 0.85\\
        \texttt{AD45-22} & $45M_\odot$ & original &$32.6M_\odot$& $22.0M_\odot$ & 0.64 & 2.71&0.03 & 528 & $28.0M_\odot$ & 0.77\\
        \texttt{AD45-25} & $45M_\odot$ & original &$32.4M_\odot$& $25.0M_\odot$ & 0.73 & 1.45&0.03 & 600 & $27.7M_\odot$ & 0.75\\
        \texttt{AD45-25-hv} & $45M_\odot$ & original &$32.4M_\odot$& $25.0M_\odot$ & 0.73 &1.45& 0.10 & 600 & $26.8M_\odot$ & 0.74\\
        \hline
        \texttt{AD20x1} & $20M_\odot$ & original &$15.1M_\odot$& --- & --- & --- & --- & 175 & $11.2M_\odot$ & 0.73\\
        \hline 
    \end{tabular}\\
    \label{tab:model}
\end{table*}

The next step is to determine the profile of the infalling matter located outside the black hole. For this, we approximate that the envelope in the progenitor stars is in a free-fall state during the collapse. To characterize the profile, we employ a solution of Oppenheimer-Snyder collapse (e.g., Ref.~\cite{Petrich:1985jko}) for our free-fall approximation because the centrifugal effect before the disk formation is minor for the collapsing matter. Then, the fluid motion in the stellar envelope during the collapse is given by
\beqn
    r_m(\tau_m) &=& \frac{1}{2} r_{m,0} \left( 1+ \cos\eta \right), \\
    \tau_m &:=& \max(\tau - \tau_{m,0},0) = \sqrt{\frac{r_{m,0}^3}{8m}} \left(\eta + \sin \eta \right),
\eeqn
where $r_m$ is the areal radius of the mass shell with the enclosed mass $m$, $r_{m,0} = r_m(\tau_m=0)$, $\tau_{m,0}$ is the starting time of the free-fall (see below), $\tau_m$ is the free-fall time of the mass shell, and $\eta$ is an auxiliary parameter. For simplicity, we assume that the matter in the envelope has zero radial velocity initially and begins to free-fall when the sound wave propagated from the center reaches the radius at
\beq
    \tau_{m,0} = \int_0^{r_{m,0}} \frac{dr}{c_\mathrm{s}(r)}.
\eeq
Then, the black-hole formation time $\tau=\tau_\mathrm{BH}$ can be estimated as
\begin{align}
    \tau_\mathrm{BH} &= \sqrt{\frac{R_\mathrm{BH,0}^3}{8M_\mathrm{BH,0}}} \left(\eta_\mathrm{BH} + \sin \eta_\mathrm{BH} \right) + \int_0^{R_\mathrm{BH,0}} \frac{dr}{c_\mathrm{s}(r)}, 
\end{align}
where $\cos \eta_\mathrm{BH} = 4 M_\mathrm{BH,0}/R_\mathrm{BH,0} - 1$ and $R_\mathrm{BH,0}$ is the areal radius of a mass shell with enclosed mass $M_\mathrm{BH,0}$. Note that the mass shell for $\tau_{m,0}>\tau_\mathrm{BH}$ does not start infalling. The radial velocity of the matter is then given approximately by
\beq
    u^r = \frac{\pa r_m}{\pa \tau} = \sqrt{\frac{2 m \left(r_{m,0} - r_m(\tau_m) \right)}{r_{m,0}r_m(\tau_m)}}.
\eeq

Since we use the spinning black-hole puncture in quasi-isotropic coordinates for the  initialization of geometric variables (see Appendix~\ref{A1}),
we need to perform coordinate transformation to quasi-isotropic coordinates $(\bar{r},\theta,\varphi)$ for consistency as 
\beq
    \bar{r} = \frac{1}{2}\left( r_m - m + \sqrt{r_m^2 - 2m r_m + a_{m}^2}\right),
\eeq
where $a_m=J(m) / m$ and we assumed the conservation of the rest mass, $m$, and angular momentum $J(m)$ along radial geodesics of infalling mass shells.
As a result, the weighted rest-mass density $\rho_*$, angular momentum density $\hat{J}_\varphi$, and radial velocity $u_{\bar{r}}$ (see Appendix~\ref{A1} for the definition of them) are given by
\beqn
    \rho_* &=& \frac{1}{4 \pi \bar{r}^2} \frac{\pa m}{\pa \bar{r}}, \\
    \hat{J}_\varphi &=& \frac{3}{8\pi \bar{r}^2} \frac{\pa J(m)}{\pa \bar{r}} \sin^2{\theta}, \\
    u_{\bar{r}} &=& \frac{r_m^2}{\bar{r}^2}\frac{\pa \bar{r}}{\pa r_m} u^r
    \nonumber \\
    &=&\frac{r_m^2}{\bar{r}\left(m + 2\bar{r} - r_m \right)}\sqrt{\frac{2 m \left(r_{m,0} - r_m \right)}{r_{m,0}r_m}},
\eeqn
while other thermodynamical quantities such as the specific enthalpy ($h$)  
and temperature ($T$) are obtained from the initial entropy of the matter assuming the adiabatic flow. In addition, we assume that the electron fraction is unchanged in the free-fall. After all the hydrodynamical quantities are set, we initialize the geometrical quantities following an initial-value formulation presented in Appendix~\ref{A1}. 

The initial data is prepared using the multigrid solver code modified based on octree-mg \cite{octreemg}, an open source multigrid library, with an octree adaptive-mesh refinement (AMR) grid. This code can provide more accurate initial data than in our previous paper~\cite{Fujibayashi2022dec}, and hence, enables us to explore the explosion energy and ejecta mass, which are sensitive to the accuracy of the gravitational field in the outer region of progenitor stars, with a better accuracy.  

In numerical computation, we cut out the outer part of the progenitor stars with $r \agt 10^5$\,km, because our simulation time is at most $\sim 20$\,s, and hence, the matter in such an outer region does not fall into the central region, i.e., it does not give any effect on the evolution of a black hole and a disk/torus. 



Table~\ref{tab:model} lists the models employed and their parameters, i.e., the initial total rest mass in the computational domain (including that of the matter transformed to the black hole), the initial mass and dimensionless spin of the black hole, the ratio of the matter angular momentum to the black-hole angular momentum, the alpha viscous parameter (see Sec.~\ref{sec:results} for the definition), the grid spacing that covers the central region as well as the mass and dimensionless spin of the black hole at the termination of each simulation. The last number for the model name denotes the initial black-hole mass.
Here, the black-hole mass is determined from the equatorial circumferential radius, $C_e$, of apparent horizons~(e.g., see Ref.~\cite{Shibata2016a}) by
\beq
M_\mathrm{BH}= {C_e \over 4\pi}. \label{eq:mbh}
\eeq
The dimensionless spin, $\chi$, is determined from the ratio of the meridian circumferential radius $C_p$ to $C_e$ using the relation between $\chi$ and $C_e/C_p$ for Kerr black holes~\cite{Shibata2016a}. We also confirm that the area of the apparent horizons, $A_\mathrm{AH}$, is written as $A_\mathrm{AH}=8\pi M_\mathrm{BH}^2(1 + \sqrt{1-\chi^2})$ for the given set of $M_\mathrm{BH}$ and $\chi$ within 0.1\% error. 


For the models with $M_\mathrm{ZAMS}=20$, $35$, and $45M_\odot$, the rest-mass of the matter located outside the black hole is $\approx 7$, 10, and $10M_\odot$ for $M_\mathrm{BH,0}=8$, $15$, and $22M_\odot$. This suggests that for the $35M_\odot$ and $45M_\odot$ models, the energy source available for the explosion is larger. For the stellar models of Ref.~\cite{Aguilera-Dena2020oct}, the stellar radius $R_* \sim 3\times 10^5$\,km depends only weakly on the stellar mass $M_*$ at the onset of the stellar core collapse. This implies that a compactness, defined by $C_*=GM_*/(c^2R_*)$, and the density at a given radius are larger for the larger values of $M_\mathrm{ZAMS}$, leading to a higher mass infall rate. This dependency is reflected in the explosion energy as discussed in Sec.~\ref{secIIIB}. It should be also mentioned that the angular momentum of the matter outside the black hole, $J_\mathrm{mat}$, is larger than that of the black hole, $J_\mathrm{BH,0}=\chi_0{M_\mathrm{BH,0}}^2$, for all the models with the original angular velocity. 


In this paper, the model with $M_\mathrm{ZAMS}=35M_\odot$ and $\alpha_\nu=0.03$ (\texttt{AD35-15}) is taken as a fiducial model. We perform additional simulations by uniformly multiplying constant factors 0.5, 0.6, 0.8, and 1.2 to the angular velocity of this fiducial model (each is referred to as \texttt{AD35-15x0.5}, \texttt{AD35-15x0.6}, \texttt{AD35-15x0.8}, and \texttt{AD35-15x1.2}). This exploration is motivated by the fact that the stellar evolution calculation is carried out assuming the spherical morphology and the results for the angular velocity profile may have a systematic uncertainty. By varying the angular velocity we explore the dependence of the ejecta mass and explosion energy on the initial angular momentum. We also perform simulations with $\alpha_\nu=0.06$ and 0.10 for the model with $M_\mathrm{ZAMS}=35M_\odot$. 

As we already mentioned, Fig.~\ref{fig1} indicates that it would be safe to choose $M_\mathrm{BH,0} \approx 8$, $15$, and $22M_\odot$ at which a disk starts forming. By performing numerical simulations, we find that it is practically possible to employ larger values of $M_\mathrm{BH,0}$, because in an early stage of the disk evolution during which the viscous timescale of the disk is shorter than its growth timescale, the matter in the disk quickly falls into the black hole. Thus, we also employ $M_\mathrm{BH,0}=9$ and $10M_\odot$ for $M_\mathrm{ZAMS}=20M_\odot$ and $M_\mathrm{BH,0}=25M_\odot$ for $M_\mathrm{ZAMS}=45M_\odot$. With these settings, the computational costs are saved because we can employ a larger grid spacing (see Sec.~\ref{sec:results}). Although the setting is different from the more reliable one (with a smaller value of $M_\mathrm{BH,0}$), it is indeed found that the results for the explosion energy and ejecta mass depend only weakly on the initial choice of $M_\mathrm{BH,0}$ if the boost of $M_\mathrm{BH,0}$ is within $\sim 15\%$. However, $M_\mathrm{BH,0}$ should not be taken to be too large. For example, for $M_\mathrm{ZAMS}=20M_\odot$ with $M_\mathrm{BH,0}=10M_\odot$, the final black-hole spin is overestimated, because a part of the high-angular-momentum matter that should form the disk in reality is incorrectly taken inside the black hole for the initial condition.

\section{Numerical results}
\label{sec:results}

\begin{figure*}
\includegraphics[width=0.32\textwidth]{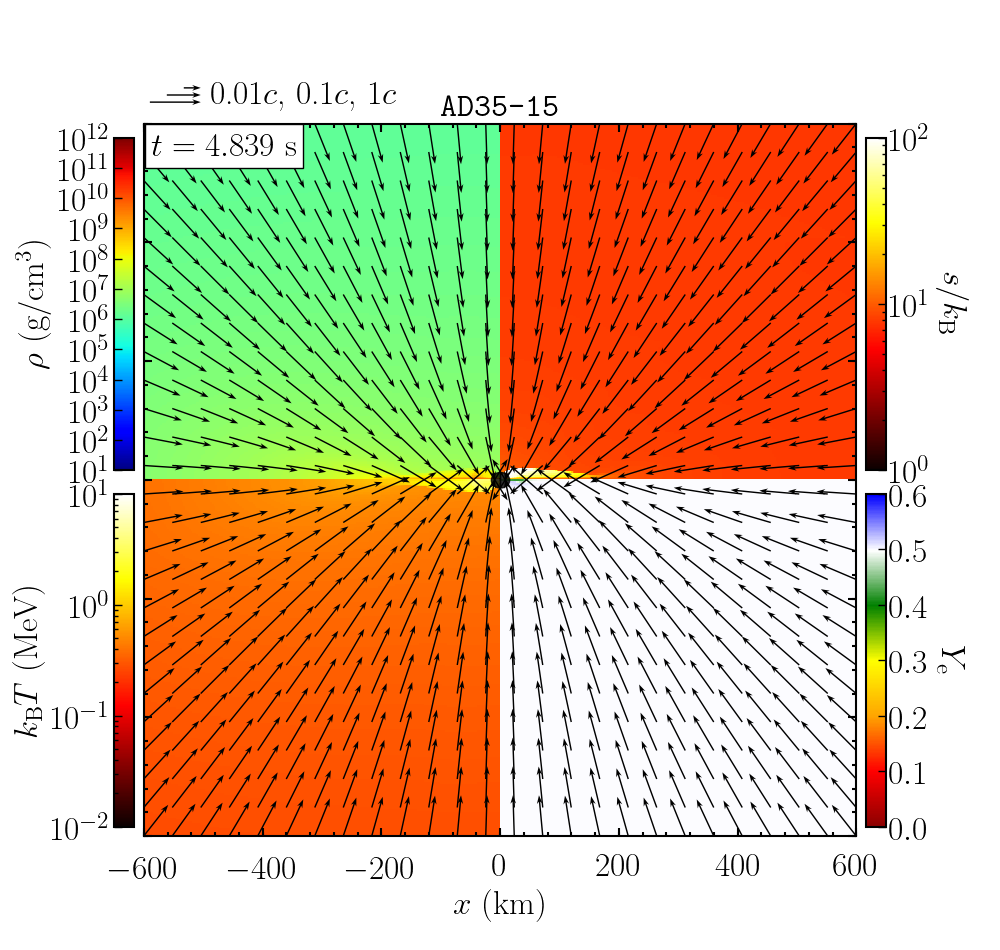}
\includegraphics[width=0.32\textwidth]{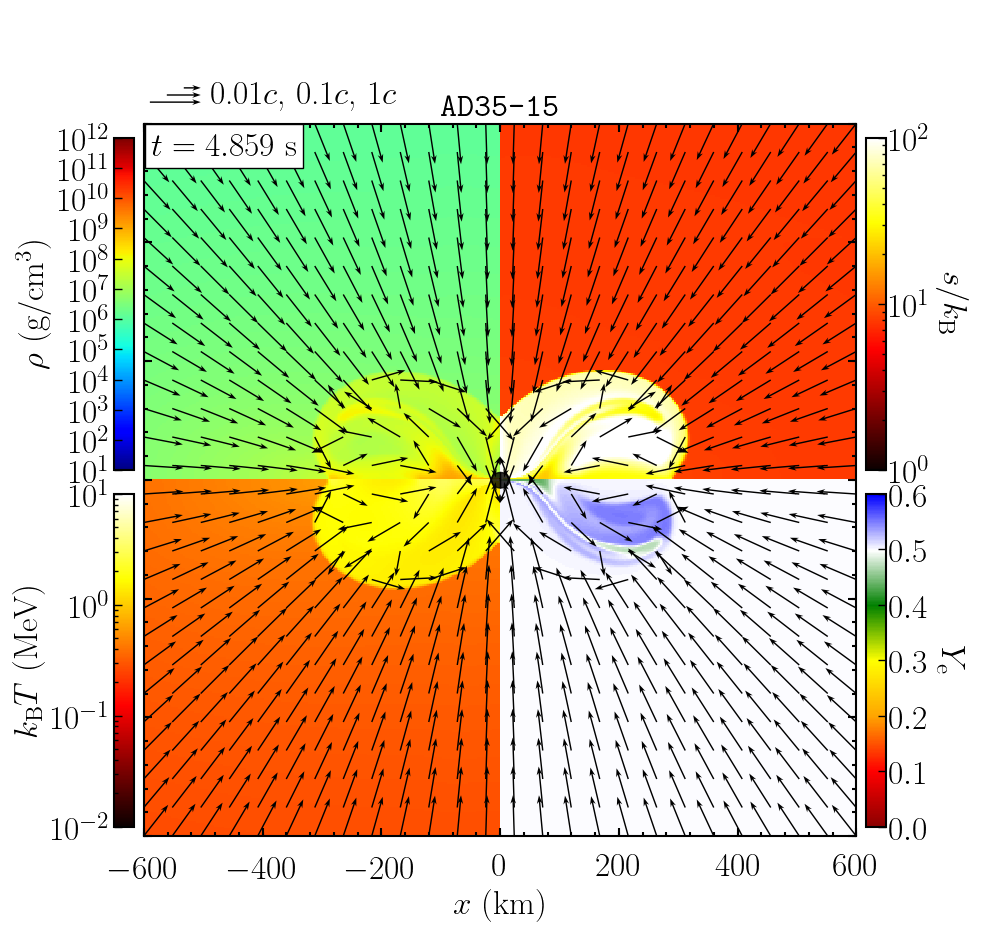}
\includegraphics[width=0.32\textwidth]{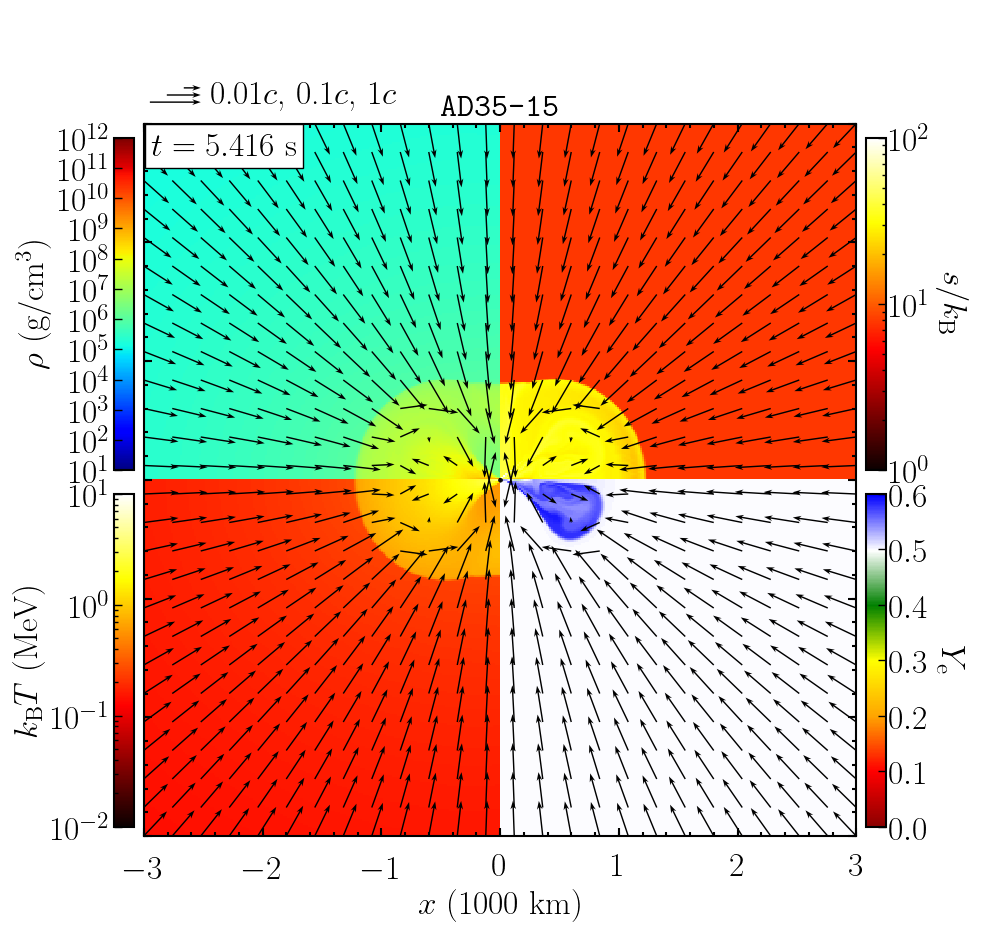}\\
\includegraphics[width=0.32\textwidth]{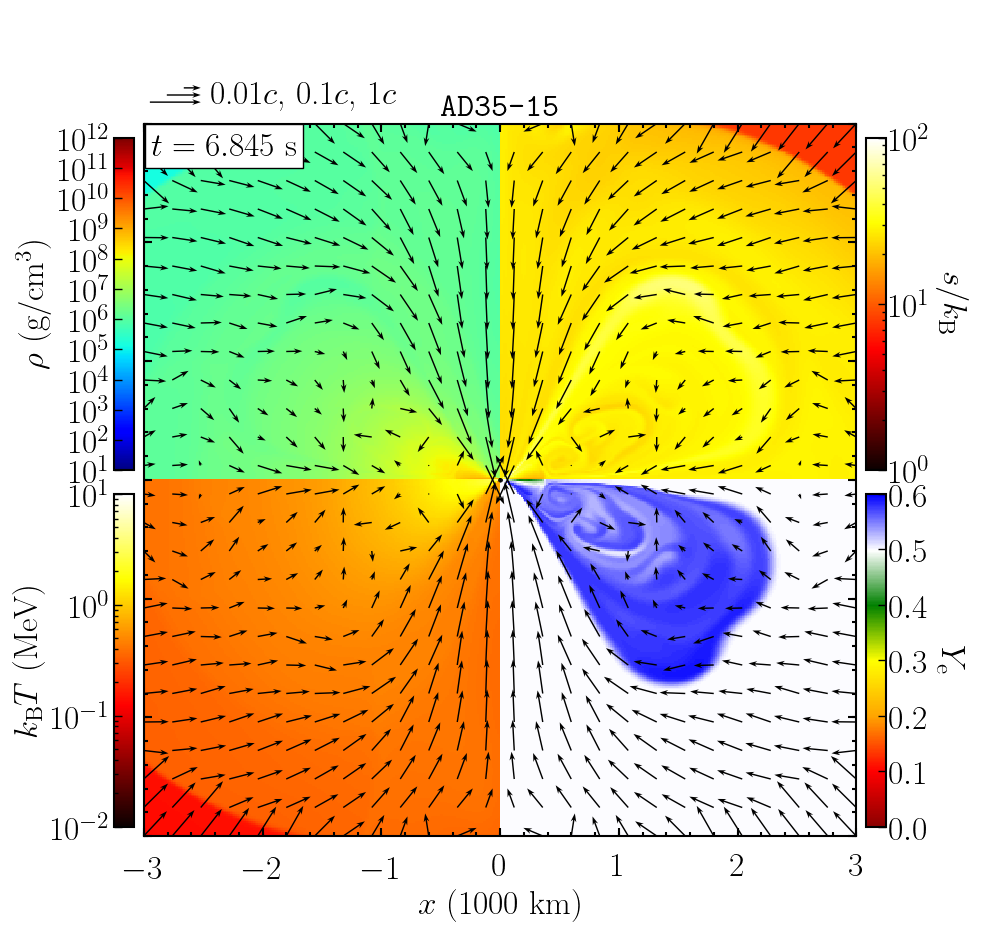}
\includegraphics[width=0.32\textwidth]{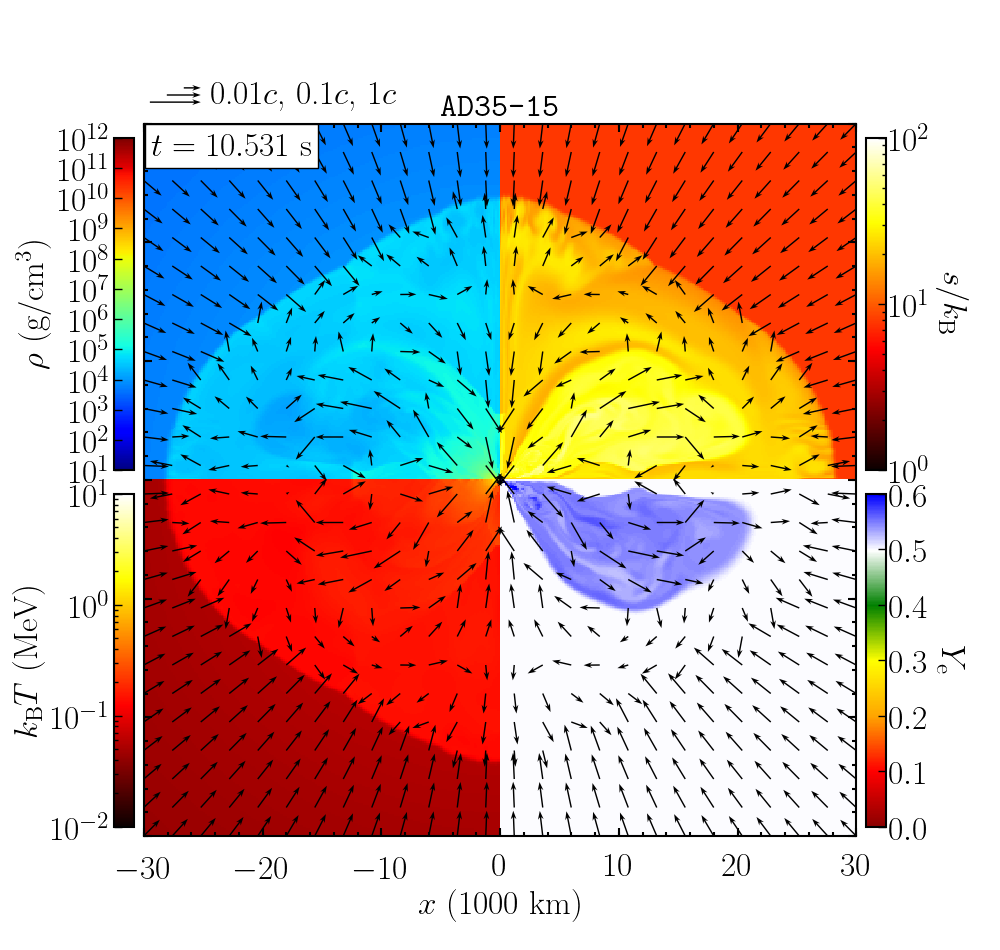}
\includegraphics[width=0.32\textwidth]{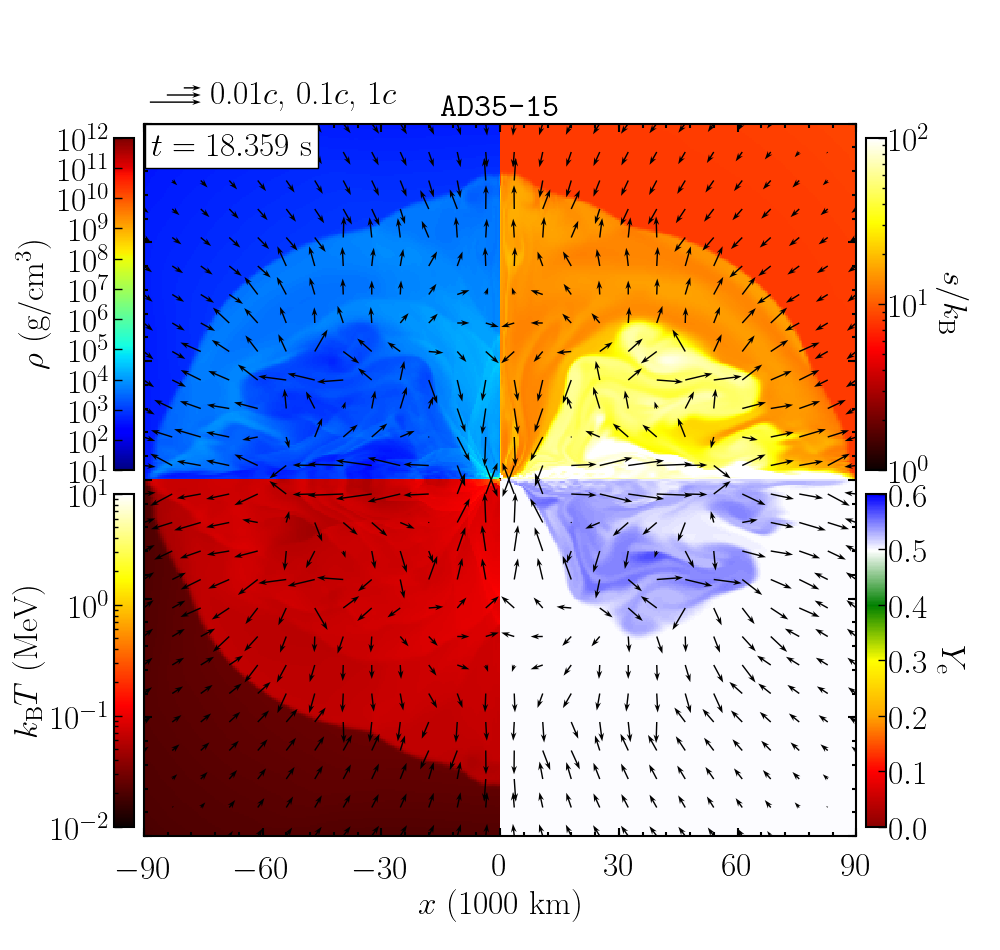}
\caption{Snapshots of the profiles for several quantities at selected time slices for model \texttt{AD35-15}. At each time, the rest-mass density (top-left), entropy per baryon (top-right), temperature (bottom-left), and electron fraction (bottom-right) are displayed.
The poloidal velocity field is depicted with arrows, the length of which is logarithmically proportional to the magnitude of the poloidal velocity. See the key shown in the top-left legend for the scale. Note that for the third to sixth panels, the regions displayed are wider than those for the first and second panels. The filled circles at the center denote the inside of apparent horizons. An animation for this model can be found in \url{https://www2.yukawa.kyoto-u.ac.jp/~sho.fujibayashi/share/AD35-15-multiscale.mp4}
}
\label{fig2}
\end{figure*}


\begin{figure*}
\includegraphics[width=0.32\textwidth]{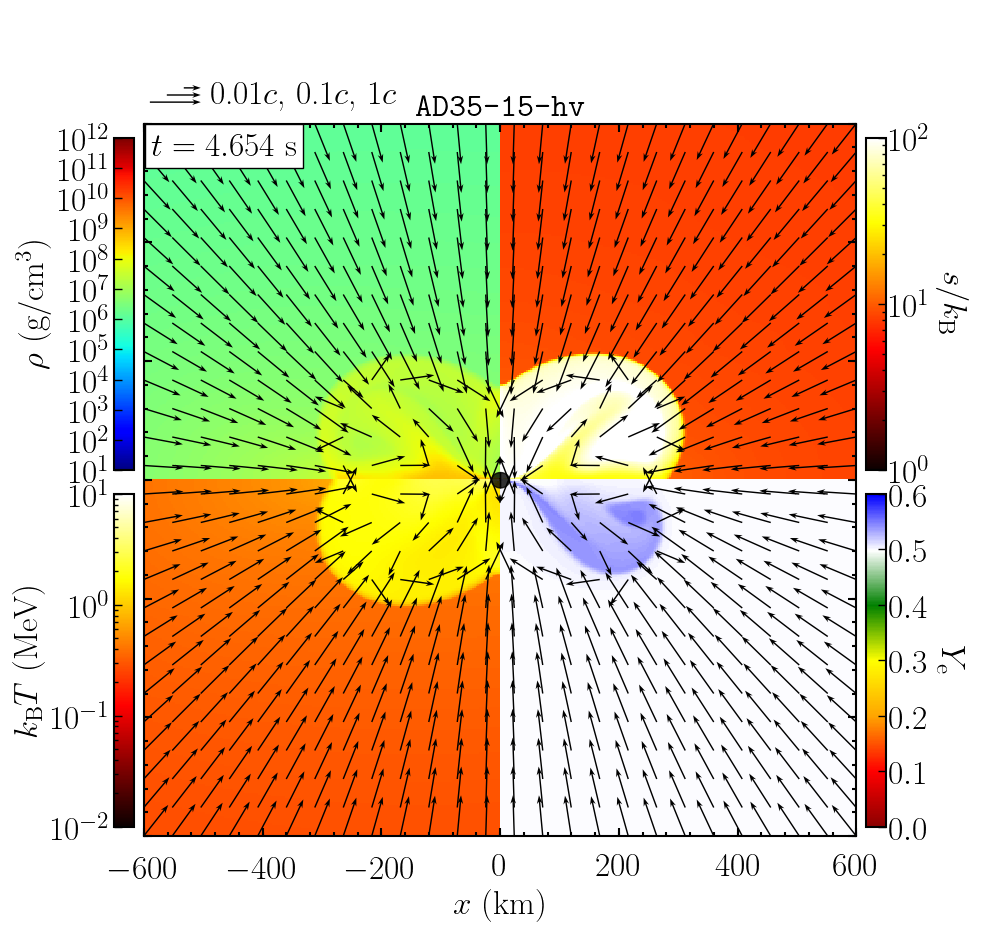}
\includegraphics[width=0.32\textwidth]{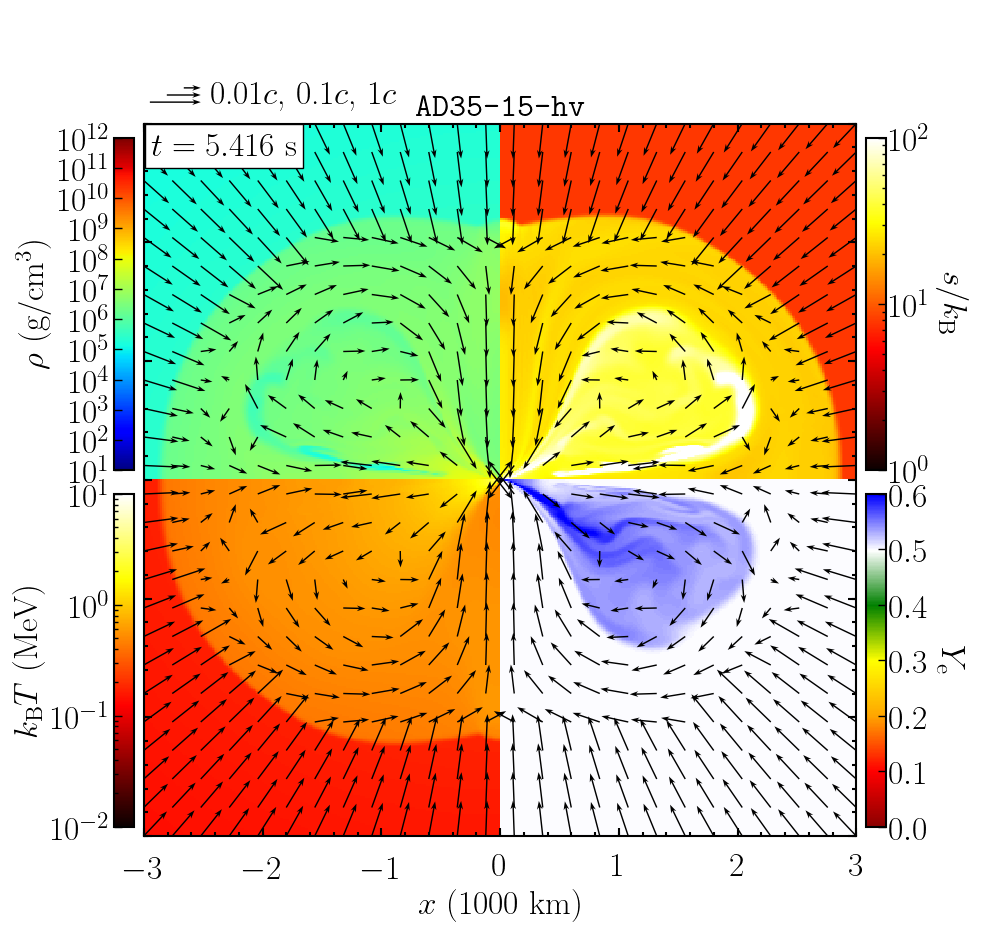}
\includegraphics[width=0.32\textwidth]{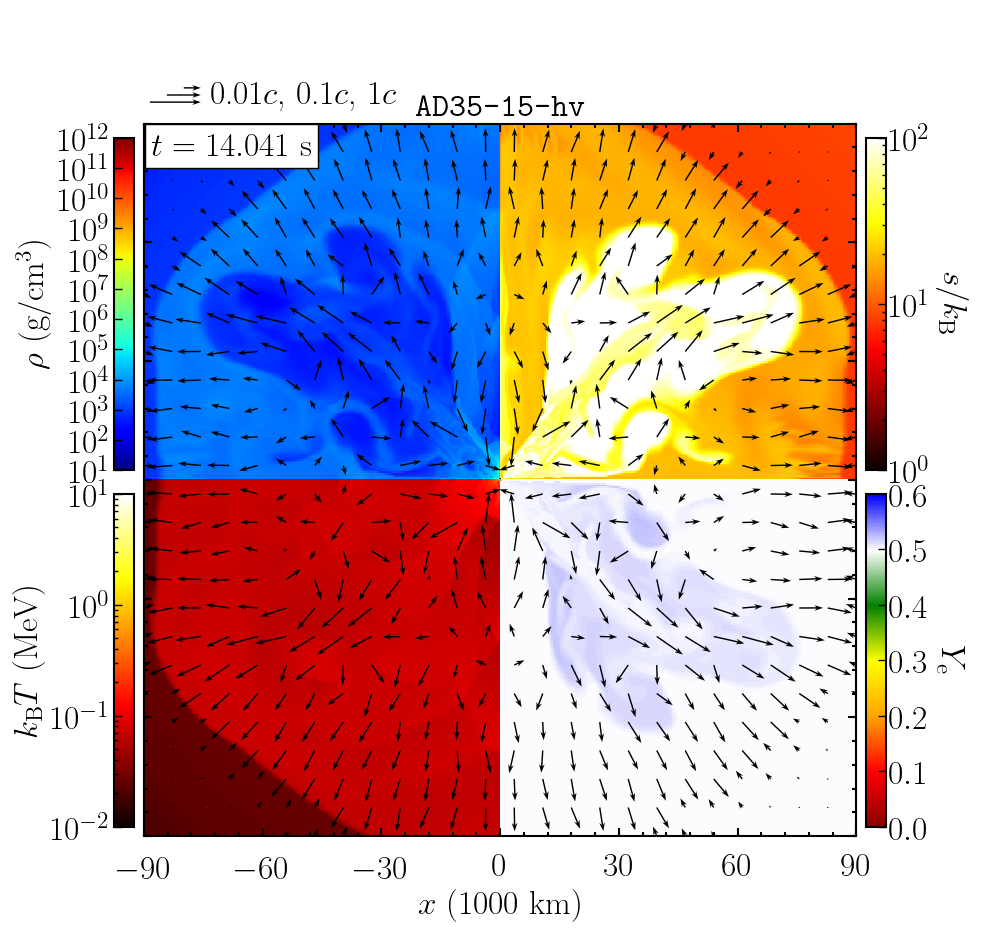}
\caption{The same as Fig.~\ref{fig2} but for larger viscosity model \texttt{AD35-15-hv}. An animation for this model can be found in \url{https://www2.yukawa.kyoto-u.ac.jp/~sho.fujibayashi/share/AD35-15-hv-multiscale.mp4}}
\label{fig2b}
\end{figure*}

\subsection{Set-up}

Numerical simulations are performed employing the same formulations as in our previous studies~\cite{Fujibayashi2020a,Fujibayashi2020b,Fujibayashi2020c}. For the viscous hydrodynamics simulation, we have to give the viscous parameter $\nu$~\cite{Fujibayashi2020a,Fujibayashi2020b,Fujibayashi2020c}. Following our previous works we write it in the form
\beq
\nu = \min(c_\mathrm{s}, 0.1c) \ell_\mathrm{tur},
\eeq
where $\ell_\mathrm{tur}:=\alpha_\nu H$ is considered as a typical eddy scale in the turbulence. To conservatively incorporate the viscous effect, we set up the upper limit ($0.1c$) for the term proportional to the sound velocity in this paper.  Following previous works, we choose $H=2GM_\mathrm{BH}/c^2$, where the black-hole mass $M_\mathrm{BH}$ is determined by Eq.~\eqref{eq:mbh} at each time (see Sec.~\ref{secII}).
This choice of $H$ is conservative because it should be much larger than $2GM_\mathrm{BH}/c^2$ in an outer region of the disk/torus. However, we will show that even with such a conservative choice, the viscous effect becomes strong enough to induce a stellar explosion. In other words, the key to the explosion is the viscous effect in an inner region of the torus. 

The simulation is performed on a two-dimensional domain of $R$ and $z$ as in our previous works~\cite{Fujibayashi2020a,Fujibayashi2020b}. For both directions, the following nonuniform grid is used for the present numerical simulation: For $x \alt 7GM_\mathrm{BH,0}/4c^2$ ($x=R$ or $z$), a uniform grid with the grid spacing, typically, of $\Delta x_0\approx 0.016GM_\mathrm{BH,0}/c^2$ is used, while outside this region, the grid spacing $\Delta x_i$ is increased uniformly as $\Delta x_{i+1}=1.01\Delta x_i$, where the subscript $i$ denotes the $i$-th grid. The black-hole horizon is always located in the uniform grid zone. 

For the fiducial model with $M_\mathrm{ZAMS}=35M_\odot$ and $\alpha_\nu=0.03$, we additionally  perform a high-resolution simulation with $\Delta x \approx 0.0135M_\mathrm{BH,0}$ to examine the numerical convergence (model \texttt{AD35-15-hi}). For this we also prepare the uniform grid for $x \alt 7GM_\mathrm{BH,0}/4c^2$ and non-uniform one with $\Delta x_{i+1}=1.01\Delta x_i$ for the outer region. The dependence of the numerical results on the grid resolution is briefly summarized in Appendix~\ref{A3}. 

Because we start from the initial data of a black hole and infalling matter, we can take a large value of $\Delta x_0$ from the beginning of the simulation. For example, for $M_\mathrm{BH,0}=15M_\odot$, $\Delta x_0$ is chosen as $360$\,m (i.e., $\Delta x_0\approx 0.016M_\mathrm{BH,0}$). If we started the same simulation from the pre-collapse star, we had to prepare a computational domain that could resolve the black-hole formation and subsequent evolution. At the formation of the black hole, its mass is $\sim 3M_\odot$, and hence, if we require the grid spacing that can resolve the black hole at birth with an accuracy as good as the present setting, we have to prepare $\Delta x_0 \approx 72$\,m. Therefore by starting the simulation from a black hole and infalling matter, we can save the computational costs significantly. 

A caution is appropriate here: For the lower grid resolutions (larger values of $\Delta x_0/M_\mathrm{BH}$), the black hole is less accurately resolved, leading to the overestimation of the black-hole mass and underestimation of the black-hole spin in our implementation~\cite{Fujibayashi2020a} (see also Appendix~\ref{A2}). This is in particular the case for model \texttt{AD20-7.8} as well as for model \texttt{AD20x1} for which the early evolution of the black hole during the stage of $M_\mathrm{BH} \approx 3M_\odot$ is less accurately computed. For other models, we choose $\Delta x_0 \leq 0.016GM_\mathrm{BH}/c^2$, with which the black hole is evolved in a good accuracy (see Appendix~\ref{A2}).

As we mentioned in Sec.~\ref{secII}, we cut out the matter for $r \agt 10^5$\,km although the original stellar surface is located at $\sim 3\times 10^5$\,km. The matter in the outer region can affect the explosion dynamics when the exploded matter interacts with it. However, the total mass of the cut-out matter is about 0.6, 1.1, and $1.3M_\odot$ for $M_\mathrm{ZAMS}=20$, 35, and $45M_\odot$~\cite{Aguilera-Dena2020oct}, and thus, they are much smaller than the ejecta mass for most of the models (see Sec.~\ref{sec:results}).  

We stop the simulation when a shock wave associated with the explosion from the disk/torus reaches the outer boundary (at $r\approx 10^5$\,km) for $M_\mathrm{BH}=35M_\odot$ and $45M_\odot$. For $M_\mathrm{BH}=20M_\odot$ for which $\Delta x_0$ is small and more computational resources are required for a long-term computation, we stopped the simulations before the explosion energy and ejecta mass saturate to save the computational time, because our main focus in this paper is the explosion property for large-mass progenitor stars.

\subsection{Explosion mechanisms}\label{secIIIA}

\begin{figure*}
\includegraphics[width=0.48\textwidth]{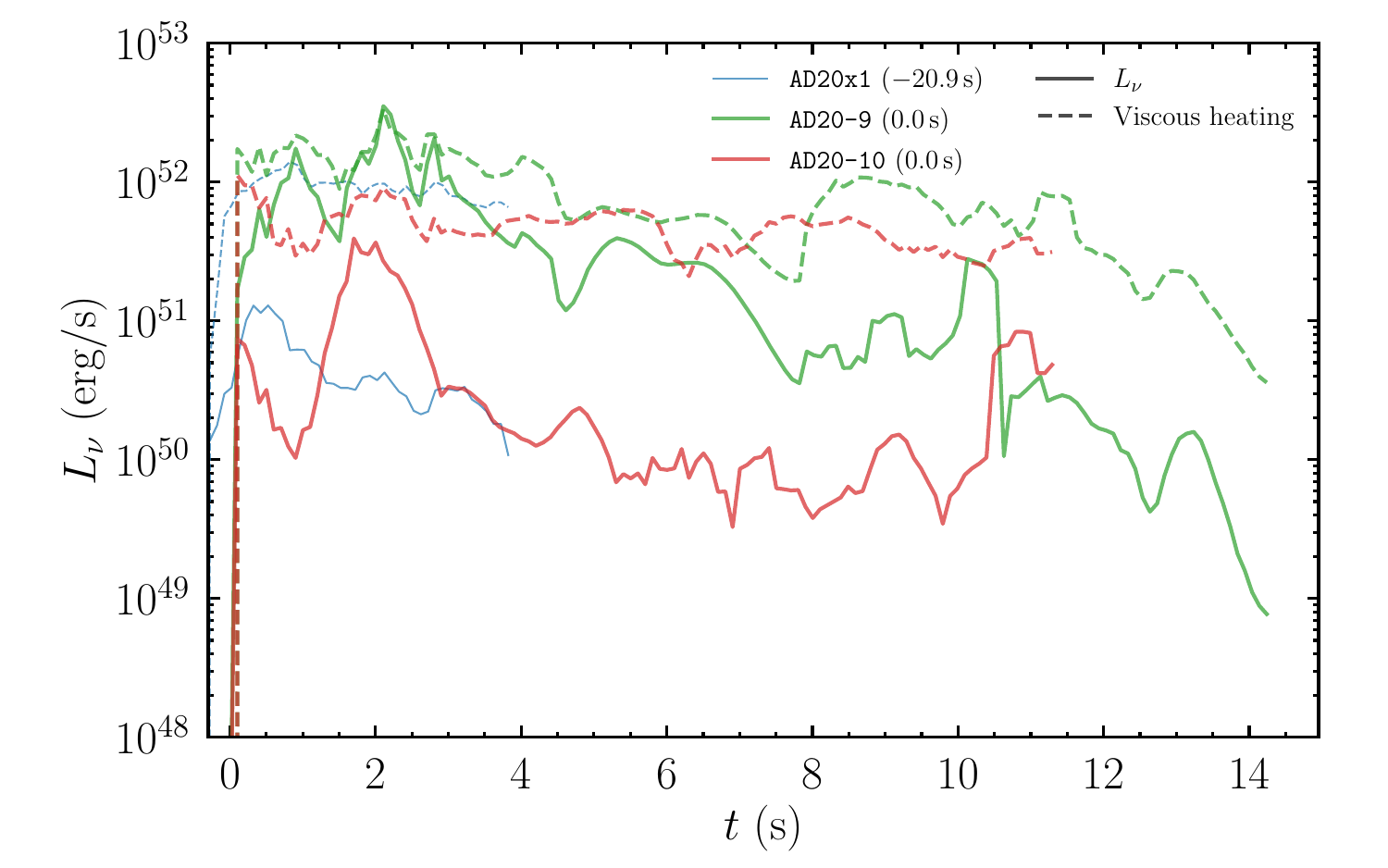}
\includegraphics[width=0.48\textwidth]{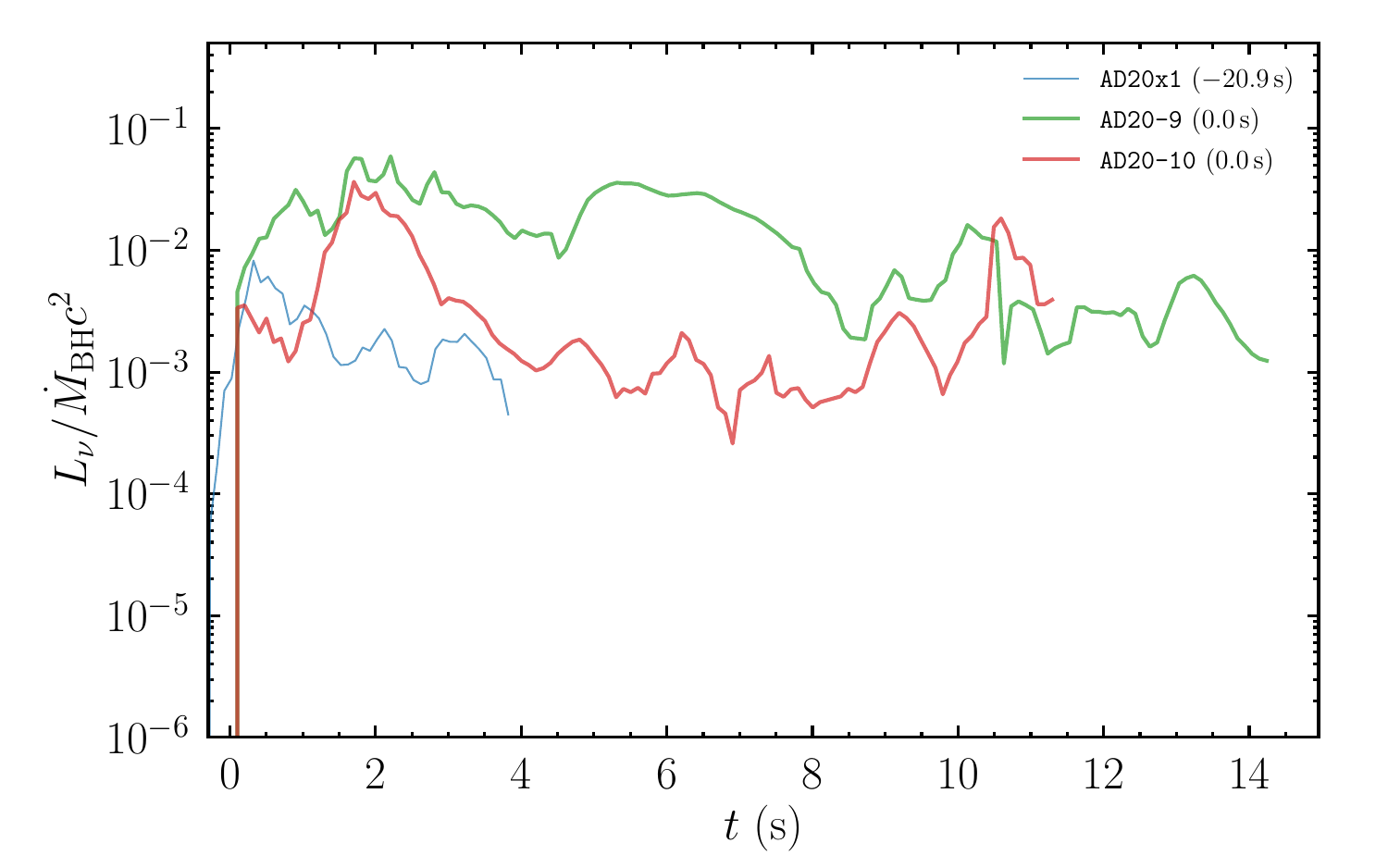}\\
\includegraphics[width=0.48\textwidth]{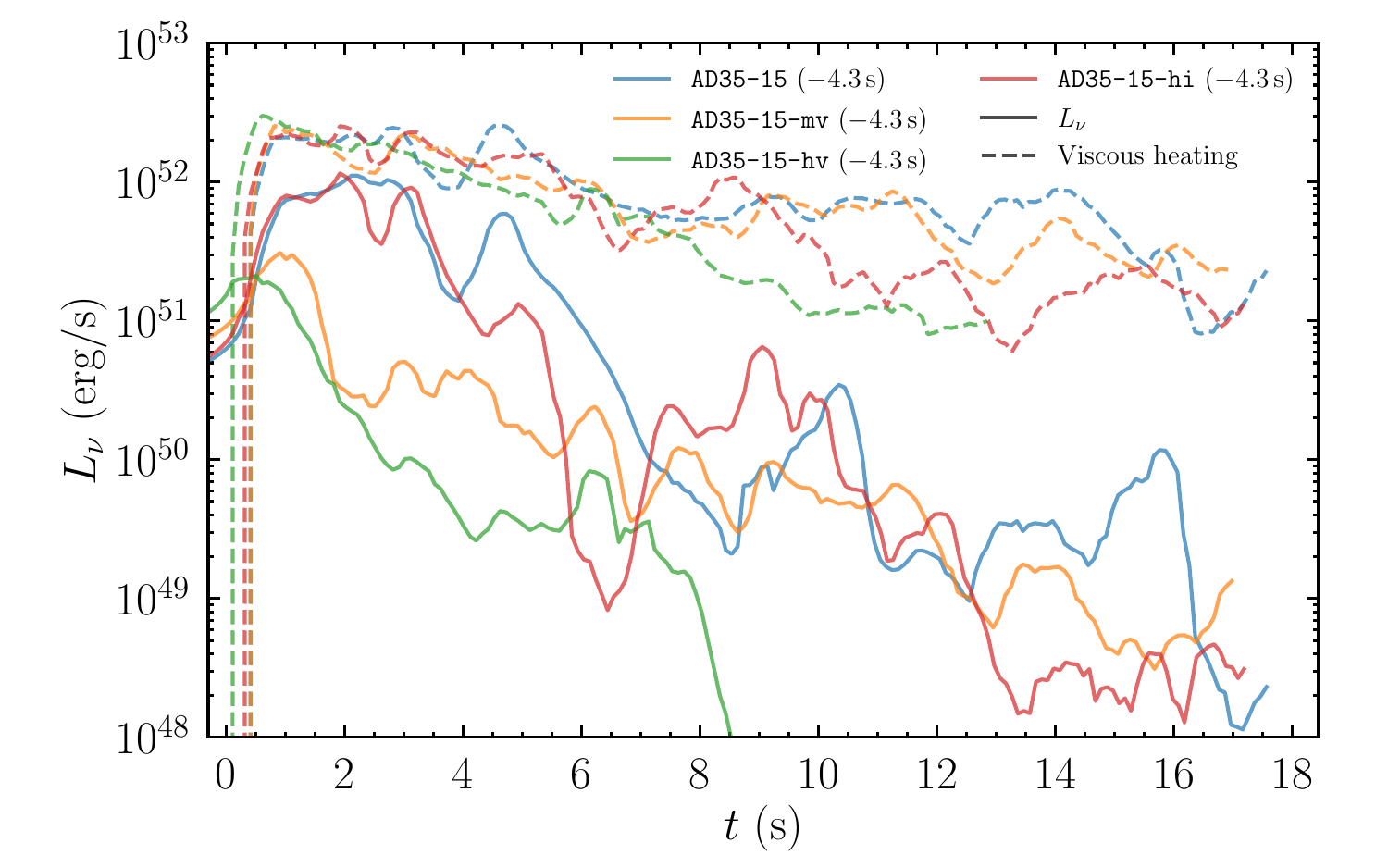}
\includegraphics[width=0.48\textwidth]{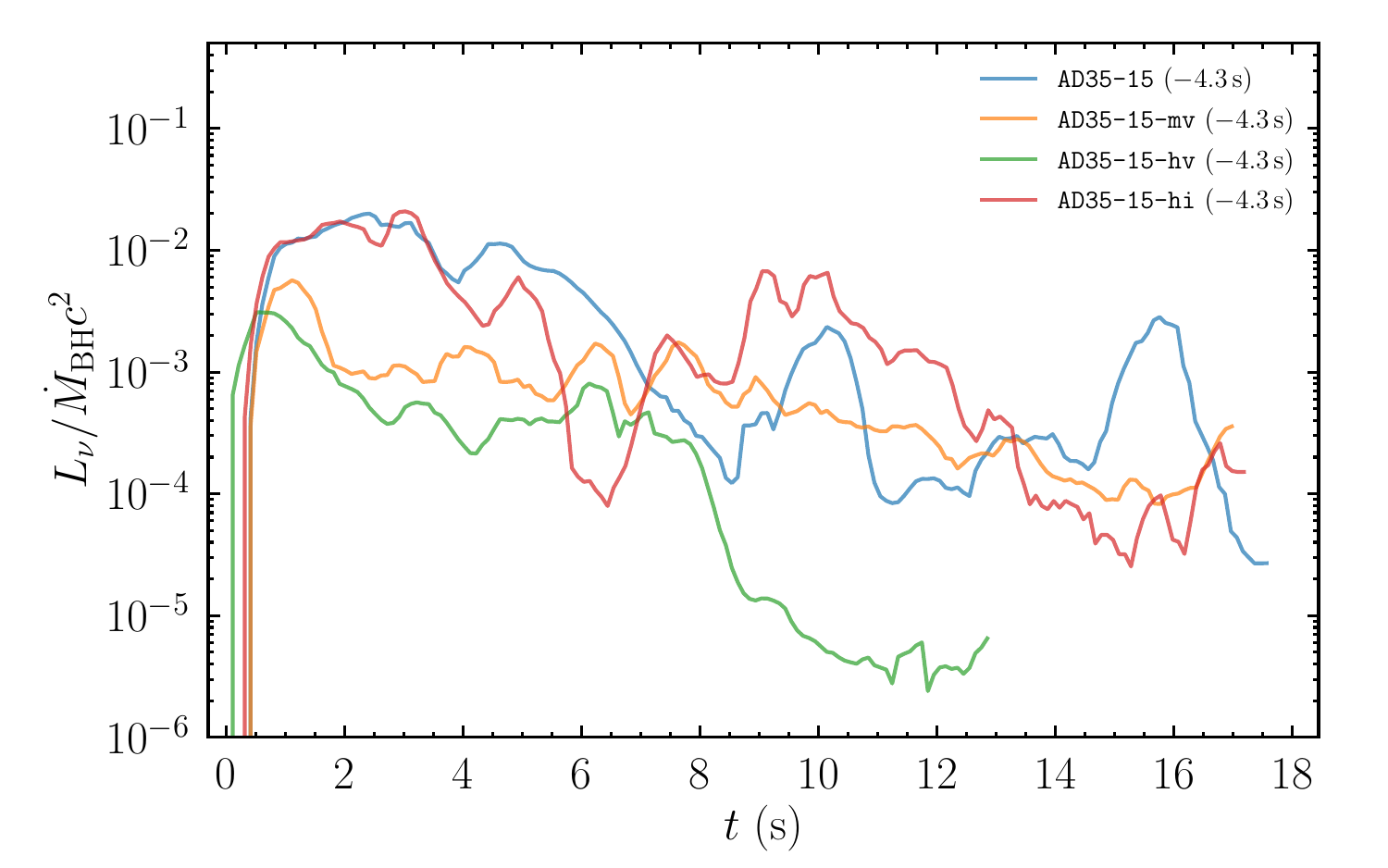}\\
\includegraphics[width=0.48\textwidth]{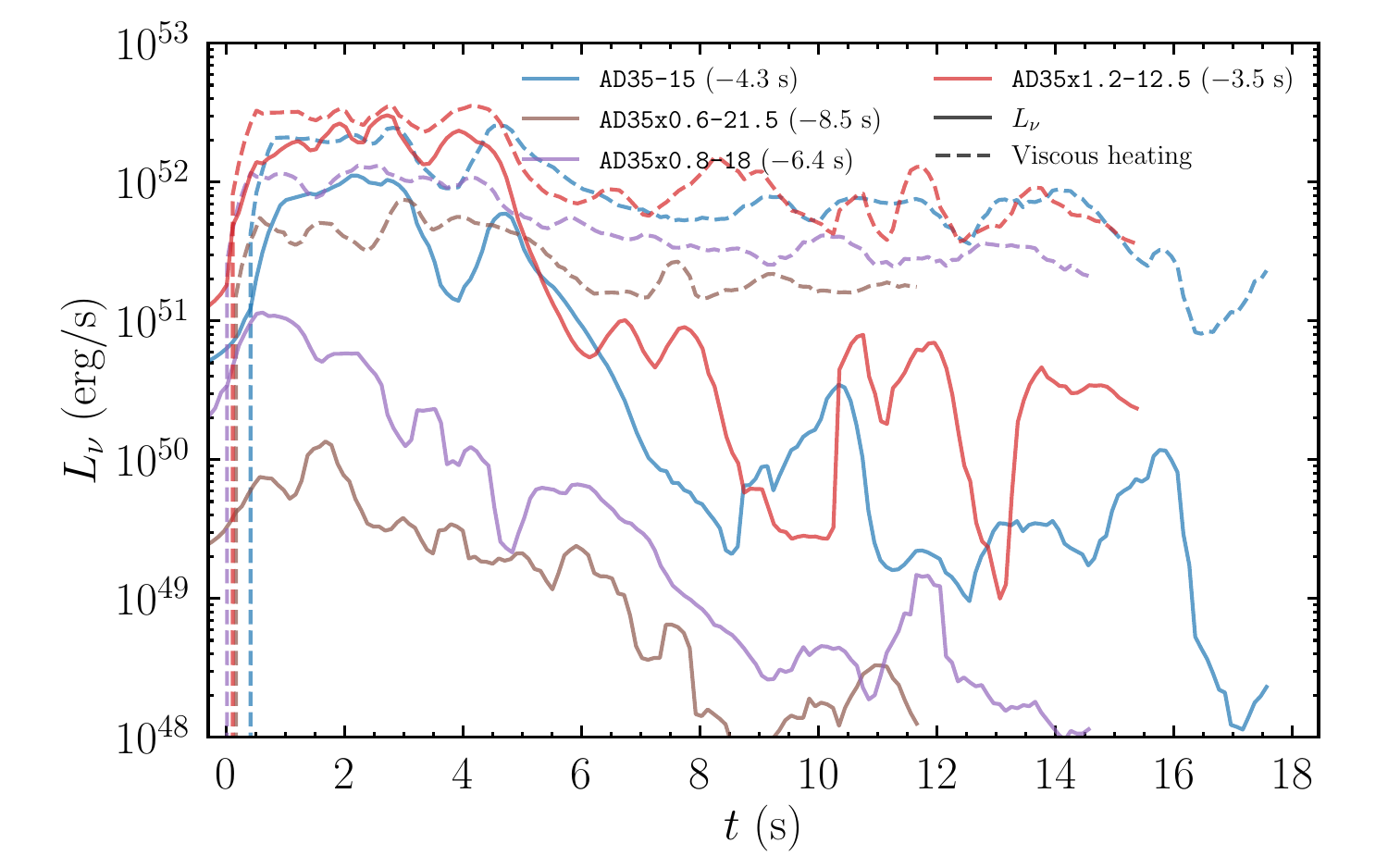}
\includegraphics[width=0.48\textwidth]{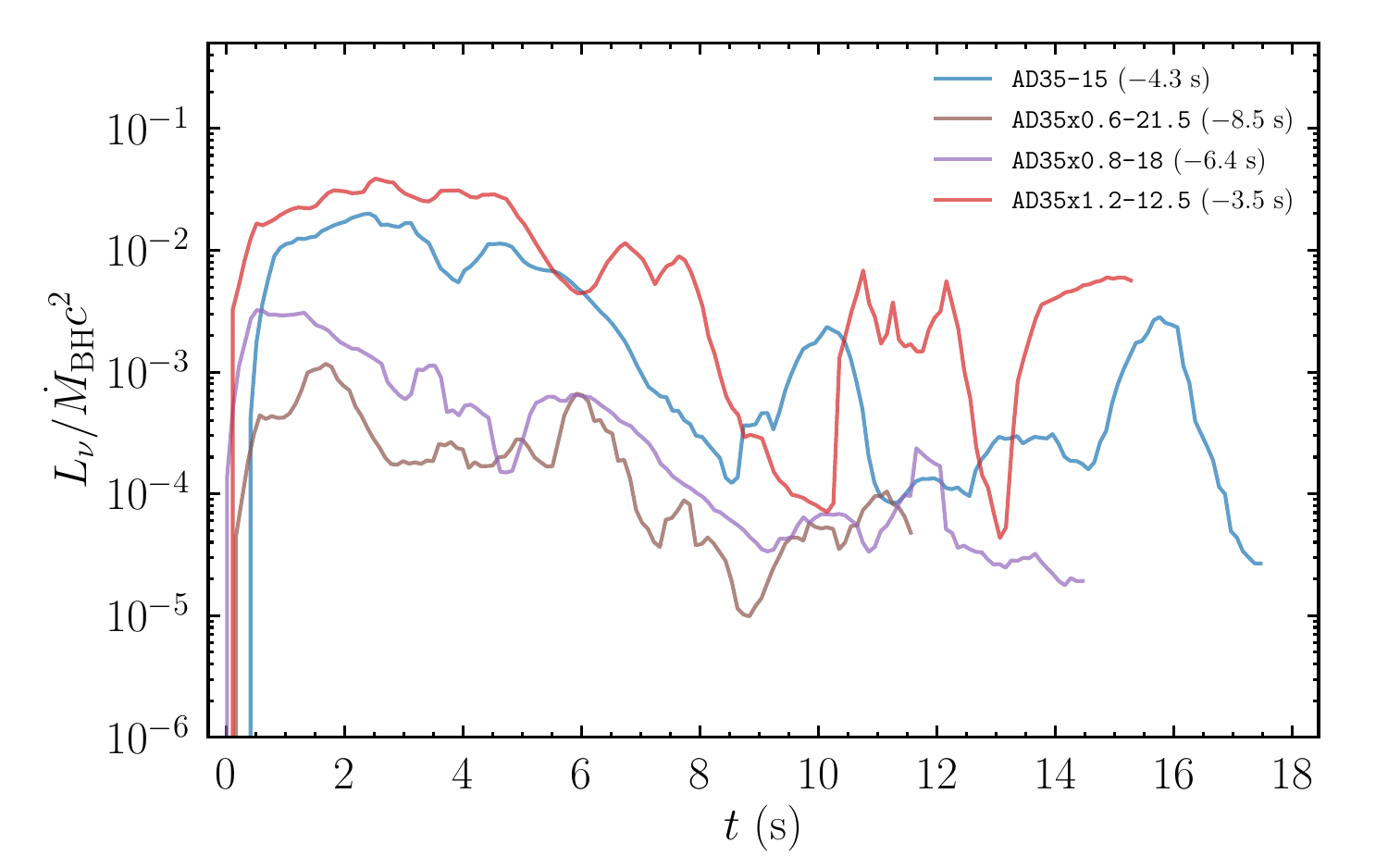}\\
\includegraphics[width=0.48\textwidth]{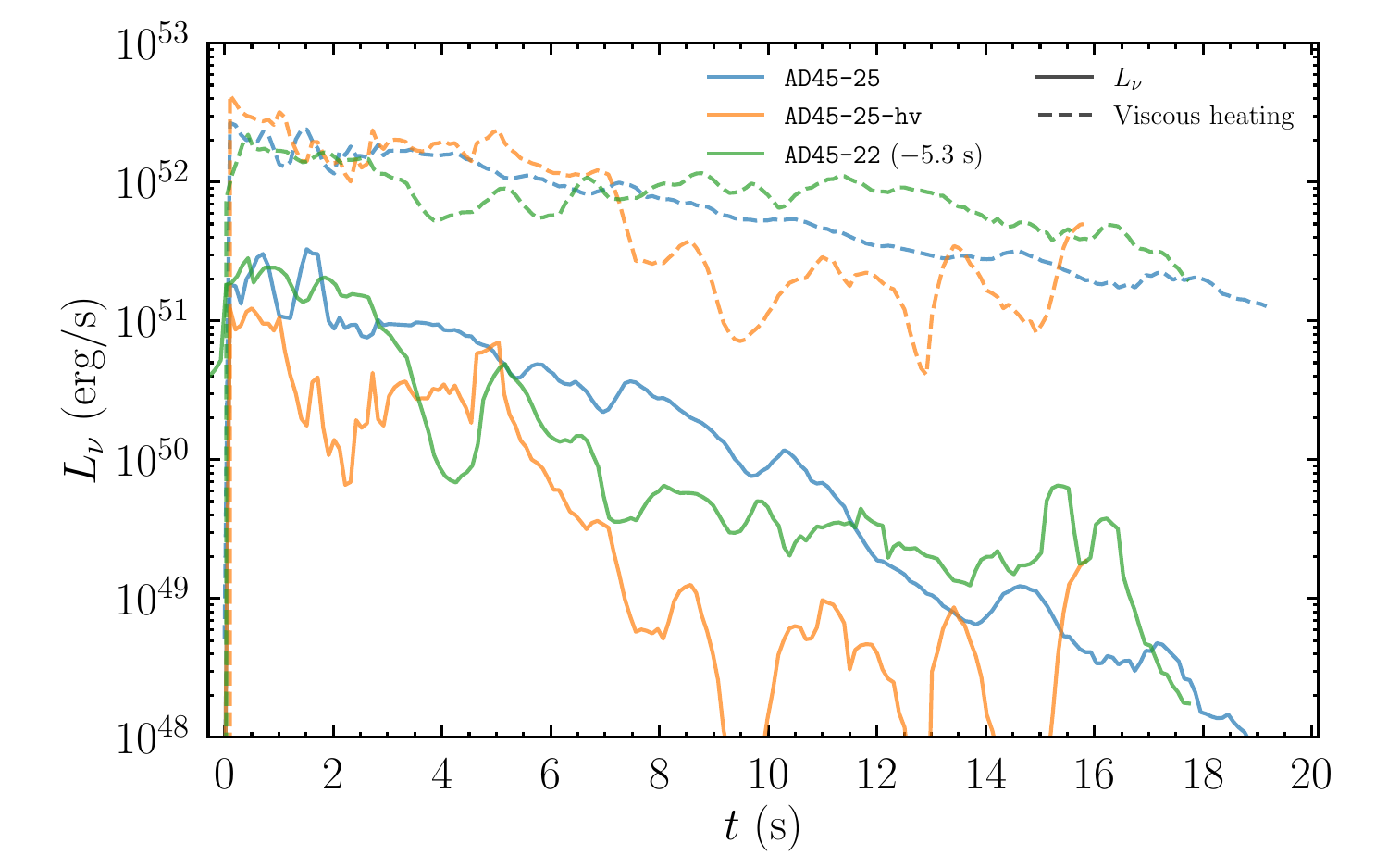}
\includegraphics[width=0.48\textwidth]{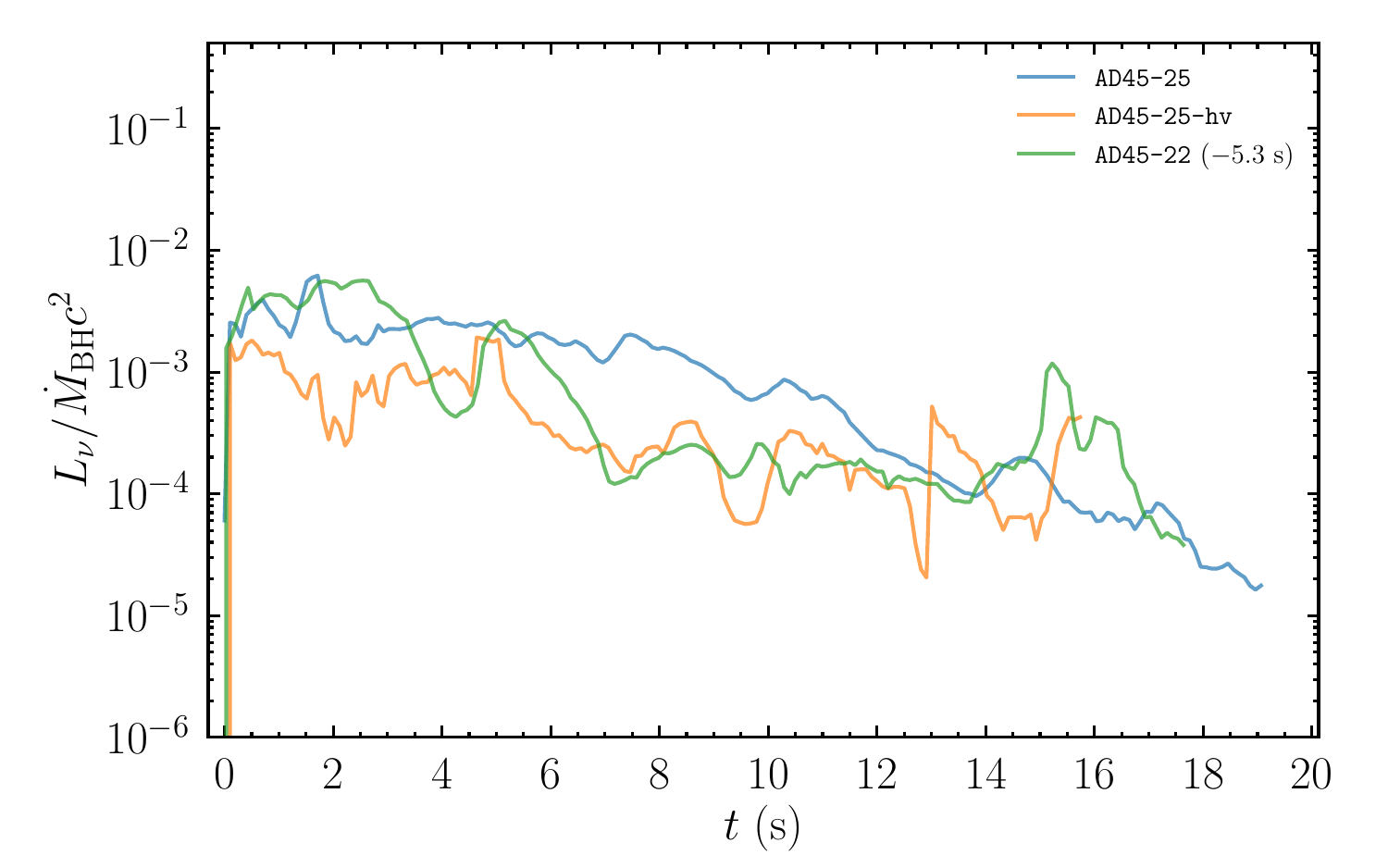}
\caption{Time evolution of the total neutrino luminosity (left) and cooling efficiency (right) for models of $M_\mathrm{ZAMS}=20M_\odot$ (top panels), $35M_\odot$ with three different values of the viscous coefficient (second top panels), $35M_\odot$ with different initial angular momentum (third top panels), and $45M_\odot$ (bottom panels). The time is shifted so that $t=0$ corresponds to the torus formation time for each model. The time offsets are shown in the legend.
}\label{fig:cool}
\end{figure*}

\subsubsection{General feature}

First, we summarize how the disk and torus are formed and evolved, leading to the eventual explosion (see Figs.~\ref{fig2} and \ref{fig2b}). As we find from Fig.~\ref{fig1}, broadly speaking, the specific angular momentum of the infalling matter increases with the enclosed mass, thus with the radius. The matter located in the inner region does not have the specific angular momentum large enough to form a disk or torus around the black hole. Thus, in an early stage of the black-hole evolution, most of the infalling matter simply falls into the black hole. During this stage, the centrifugal force of the infalling matter does not play an important role. Subsequently, the matter with sufficiently large specific angular momentum starts forming a geometrically thin disk (see the first panel of Fig.~\ref{fig2}).  After the formation of the disk, a strong shear layer is established between the infalling matter and the shock surface outside the disk. Thus, viscous heating efficiently generates the thermal energy. Also, shock dissipation efficiently proceeds around the shock surface. By these heating mechanisms, the disk subsequently becomes geometrically thick, leading to the formation of a torus (see the second panel of Fig.~\ref{fig2}). 

After its formation, the torus gradually grows due to the continuous matter infall, while the black hole grows due to the matter infall primarily from the polar region. During the evolution of the torus, the kinetic energy of the infalling matter is dissipated around the shock surface just outside the torus, which increases the temperature and entropy per baryon of the torus (see the second and third panels of Fig.~\ref{fig2} and the first panel of Fig.~\ref{fig2b}).  Since the shock surface is non-spherical while the matter infall proceeds nearly spherically, the shear layer is also formed, enhancing the viscous heating. The oblique shocks formed around the shock surface play a role in enhancing the matter infall onto the black hole and inner region of the torus from the polar region. This enhances the efficiency of the viscous heating in the inner region. 

In the early stage of the torus evolution, the ram pressure of the infalling matter is too high to induce an outflow from the torus. In addition, the neutrino cooling  suppresses the viscous heating effect. However, the ram pressure of the infalling matter continuously decreases because of the decrease in its density, and also, the neutrino cooling efficiency becomes lower in a later stage (see below for more details). As a result, the thermal pressure of the torus generated by the viscous and shock heating eventually exceeds the ram pressure. Then, an outflow from the torus sets in, inducing the explosion of the entire star (see the fourth, fifth, and sixth panels of Fig.~\ref{fig2} and the second and third panels of Fig.~\ref{fig2b}). 

The viscous heating as well as the shock dissipation are most efficient around the shock surface in the vicinity of the torus. Thus, the outward motion of the outflow is initially induced along the torus surface. The matter of the outward motion has high entropy per baryon, and thus, the outward motion accompanies convective motion, which redistributes the thermal energy to a wide region. Thus, although the matter initially moves toward a particular direction, subsequent motion becomes quasi-isotropic, and the explosion occurs in a nearly spherical way. 

Although the viscous and shock heating are universally the explosion sources, the efficiency of the heating and evolution process of the torus depend on the neutrino cooling (see Fig.~\ref{fig:cool}). In the presence of an efficient cooling by neutrinos, the torus relaxes to a neutrino-dominated-accretion-flow (NDAF) state. On the other hand, if the neutrino cooling is not efficient, the explosion takes place in the absence of the NDAF state and the explosion sets in earlier. For example, for model \texttt{AD35-15} for which the NDAF stage is present the explosion sets in at $t \sim 7$\,s while for model \texttt{AD35-15-hv} for which the NDAF stage is absent the explosion set is at $t\sim 5$\,s (compare Figs.~\ref{fig2} and \ref{fig2b}).

Even after the onset of the explosion, the matter infall continues for at least several seconds near the rotational axis, around which the matter with small specific angular momentum continuously falls onto the black hole and the inner region of the torus. This matter infall to the torus contributes to the efficient viscous and shock heating, sustaining the explosion.

\subsubsection{Dependence of the progenitor mass}

As mentioned in Sec.~\ref{secII}, more massive progenitor stars are more compact and thus have higher mass-infall rates, which are advantageous for generating more thermal energy (see below). By contrast, the neutrino luminosity tends to be smaller for more massive progenitor stars at the torus formation (compare the models with original rotation profiles \texttt{AD20-9}, \texttt{AD35-15}, and \texttt{AD45-25}: see left panels of Fig.~\ref{fig:cool}). This is due to the larger radius of the innermost stable circular orbit around the black hole for more massive models. That is, for more massive models, which form more massive black holes, the density and temperature of the torus are lower~\cite{Fujibayashi2020b}, and the neutrino luminosity is also lower. Consequently, the thermal energy generated by the viscous heating is efficiently used for the explosion of the system. Indeed the right panel of Fig.~\ref{fig:cool} shows that the neutrino cooling efficiency defined by $L_\nu/\dot M_\mathrm{BH}c^2$ is lower for more massive  progenitor models. This results in a shorter (or no) NDAF phase, leading to a quick explosion. The lower neutrino cooling efficiency, in addition to the higher mass-infall rate, is advantageous for large explosion energy (see Sec.~\ref{secIIIB}). This situation is in contrast to the usual core-collapse supernova explosion, in which higher neutrino luminosity of proto-neutron stars is advantageous for an earlier explosion~(e.g., Ref.~\cite{Janka2012a}).

For the fixed viscous parameter $\alpha_\nu=0.03$, $M_\mathrm{ZAMS}=20$ and $35M_\odot$ models (\texttt{AD20-9} and \texttt{AD35-15}) have high neutrino cooling efficiency appreciably exceeds 0.01 (see Fig.~\ref{fig:cool}), and have a NDAF phase. As a result, the explosion for these models is delayed after the torus formation. By contrast, no NDAF phase is found for $45M_\odot$ models (\texttt{AD45-22} and \texttt{AD45-25}), which drive the explosion shortly after the torus formation. We note that the presence or absence of the NDAF phase depends not only on the progenitor stars but also on the viscous coefficient and the initial angular momentum of the progenitor star, as discussed in the following subsections.

\subsubsection{Dependence on the viscous coefficient}

For the $35M_\odot$ progenitor, we perform three simulations varying the viscous coefficient and find that the evolution of the system depends qualitatively on the magnitude of $\alpha_\nu$. For large values of $\alpha_\nu$, i.e., 0.06 and 0.10, the evolution toward the explosion is the qualitatively same as those for the $45M_\odot$ models: The explosion sets in in a relatively short timescale after the formation of the torus with no NDAF phase (cf. Fig.~\ref{fig2b}). By contrast, for $\alpha_\nu=0.03$, the explosion is delayed because the neutrino cooling efficiency is sufficiently high to suppress the outward motion of the matter by the viscous and shock heating in the early evolution stage of the torus. For this model, the explosion is started only when the mass infalling rate is sufficiently low. This difference results from the stronger effects of the viscous heating and angular momentum transport for the larger viscosity, by which the torus expands more rapidly, reducing the neutrino cooling efficiency in an early stage.

\subsubsection{Dependence on the initial angular momentum}

The dependence of the evolution process of the system on the initial angular momentum is explored for the models of $M_\mathrm{ZAMS}=35M_\odot$ with a fixed value of $\alpha_\nu(=0.03)$. For our models, a disk and/or a torus surrounding a black hole is always formed, but their mass depends strongly on the initial angular momentum: For larger initial angular momentum, it is larger and, as a result, the explosion can be more energetic and mass ejection is more enhanced (see Sec.~\ref{secIIIB}). 

Models \texttt{AD35-15} and \texttt{AD35x1.2-12.5} achieve a high neutrino cooling efficiency and NDAF phase after the formation of tori (see Fig.~\ref{fig:cool}). By contrast models \texttt{AD35x0.6-21.5} and \texttt{AD35x0.8-18.0} do not achieve the NDAF phase. This illustrates that larger angular momentum stars are more subject to the NDAF phase after the formation of a torus around a black hole.

For a model with sufficiently reduced angular momentum (\texttt{AD35x0.5-21.5}), the disk is too sparse and low-mass ($\alt 0.5M_\odot$) to find explosion in our simulation time. In this case, the geometrically-thick torus formation is not also found in the simulation time. Even for this case, however, a low-mass disk may be a source of a transient at a  very late stage, i.e., $t \gg 10$\,s: As discussed in Ref.~\cite{Kashiyama:2015qfa}, in this case, the final configuration is likely to be a black hole surrounded only by a low-mass low-compactness disk, which could be evolved by a viscous hydrodynamics effect (resulting from magnetohydrodynamics turbulence) leading to mass ejection. If this happens, a blue, rapidly varying optical transient may be generated after long-term evolution of the accretion disk formed in late time~\cite{Kashiyama:2015qfa}.

\begin{figure}
\includegraphics[width=0.47\textwidth]{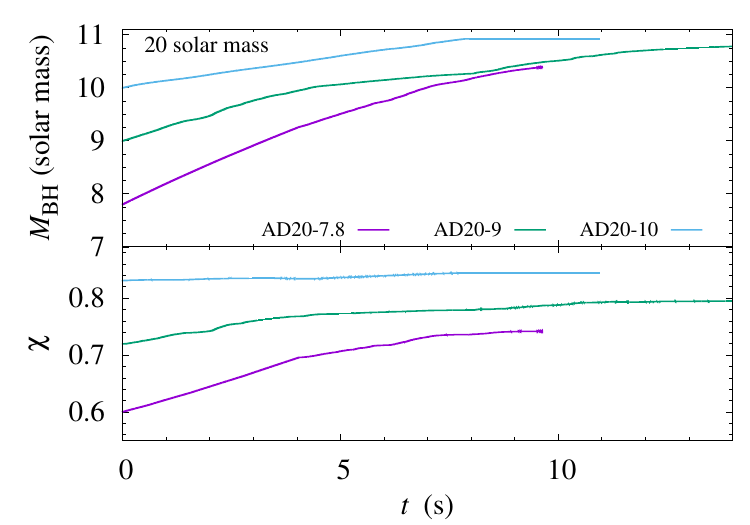}\\
\includegraphics[width=0.47\textwidth]{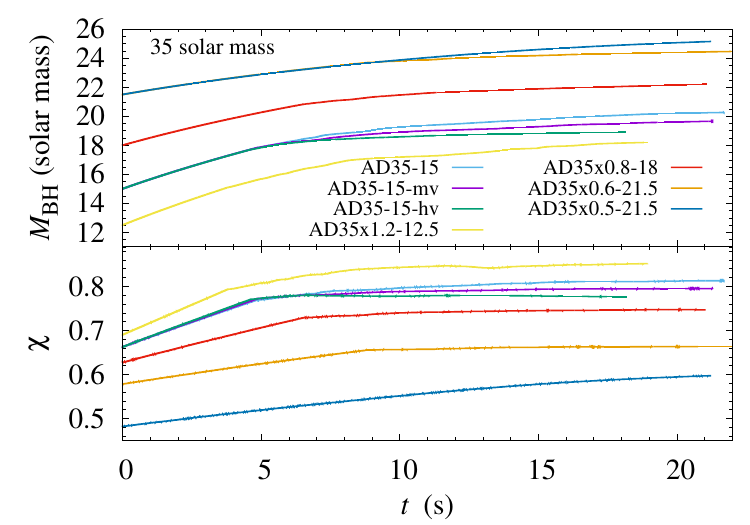}\\
\includegraphics[width=0.47\textwidth]{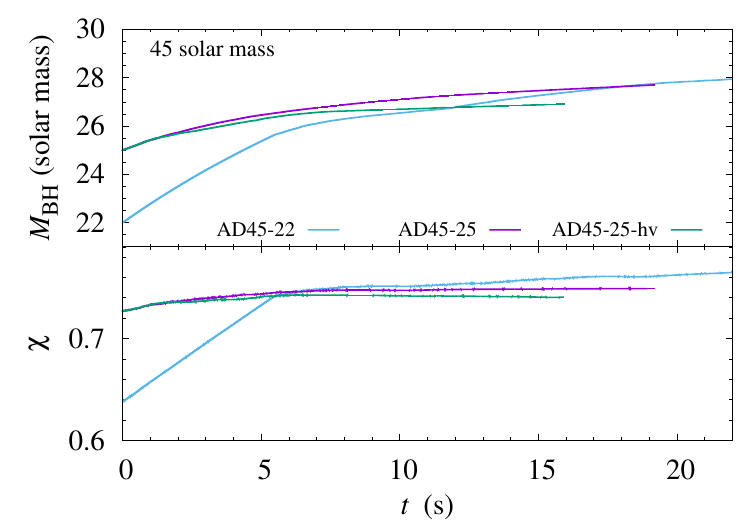}
\caption{Time evolution of the mass and dimensionless spin of the black holes for models of $M_\mathrm{ZAMS}=20M_\odot$ (upper panels), $35M_\odot$ (middle panels), and  $45M_\odot$ (lower panels). Note that for model \texttt{AD20-10}, we stopped the evolution of the gravitational field at $t \approx 8$\,s, and thus, the actual final black-hole mass may be larger. 
\label{fig:BH}
}
\end{figure}

\subsection{Evolution of black holes}
Figure~\ref{fig:BH} shows the evolution of the mass and dimensionless spin of the black holes for all the models studied in this paper. Note that for model \texttt{AD20-10}, we stopped the evolution of the gravitational field at $t \approx 8$\,s to save computational time because the total mass of the matter in the computational region was smaller than 10\% of the black-hole mass, and moreover, model \texttt{AD20-9} is our main model for $M_\mathrm{ZAMS}=20M_\odot$. 
Both the mass and dimensionless spin increase steeply prior to the onset of the explosion, but after that, they relax toward final values. The final black-hole mass is 50--60\% of $M_\mathrm{ZAMS}$; large-mass black holes such as observed by gravitational-wave observations~\cite{Abbott2021apr,Abbott2021nov} are naturally formed from the progenitor models of Ref.~\cite{Aguilera-Dena2020oct}. For the models with larger values of $\alpha_\nu$, the final mass and dimensionless spin of the black hole are slightly smaller, because higher viscous heating efficiency as well as viscous angular momentum transport enhances the mass ejection while preventing the matter infall onto the black hole. However the dependence on $\alpha_\nu$ is not very strong; the mass and dimensionless spin decrease by $\sim 1M_\odot$ and 0.03, respectively, for the change of $\alpha_\nu$ from 0.03 to 0.1. 

Accompanied with the formation of a massive disk/torus around a black hole, the black-hole spin is naturally increased. For all the models with no modification of the initial angular momentum, the dimensionless spin of the black holes is $\sim 0.75$--0.85 at the termination of the numerical simulation (cf.~Table~\ref{tab:model}). The high spin is advantageous for efficiently converting the released gravitational potential energy to the thermal energy. 

For smaller and larger initial angular momentum models with $M_\mathrm{ZAMS}=35M_\odot$, the resulting final value of the dimensionless spin of the black hole, $\chi_\mathrm{f}$, is smaller and larger, respectively, while the final black-hole mass is larger and smaller, respectively. However, $\chi_\mathrm{f}$ varies only $\pm 0.05$ for the change of the initial angular momentum by $\pm 20\%$ (compare the results for models \texttt{AD35x0.8-18}, \texttt{AD35-15}, and \texttt{AD35x1.2-12.5}). Thus, the final black hole spin is likely to be fairly high as long as a disk/torus with a few $M_\odot$ is formed around the black hole. By contrast, for model \texttt{AD35x0.5-21.5}, for which a substantial amount of the infalling matter falls into the black hole, the final value of $\chi$ is much smaller than those of the other $35M_\odot$ models, while the final mass is much larger than others. 

Models \texttt{AD45-22} and \texttt{AD45-25} started the simulations from different black-hole mass. However, the final mass and dimensional spin for these models have similar values. This appears to be also the case for models \texttt{AD20-7.8} and \texttt{AD20-9}. These results indicate that in the early stage of the disk evolution, a substantial fraction of the matter in the disk quickly falls into the black hole by the viscous effect, and the simulation may be started from a black-hole mass which is slightly larger than those predicted from Fig.~\ref{fig1}. 

\begin{figure*}[th]
\includegraphics[width=0.48\textwidth]{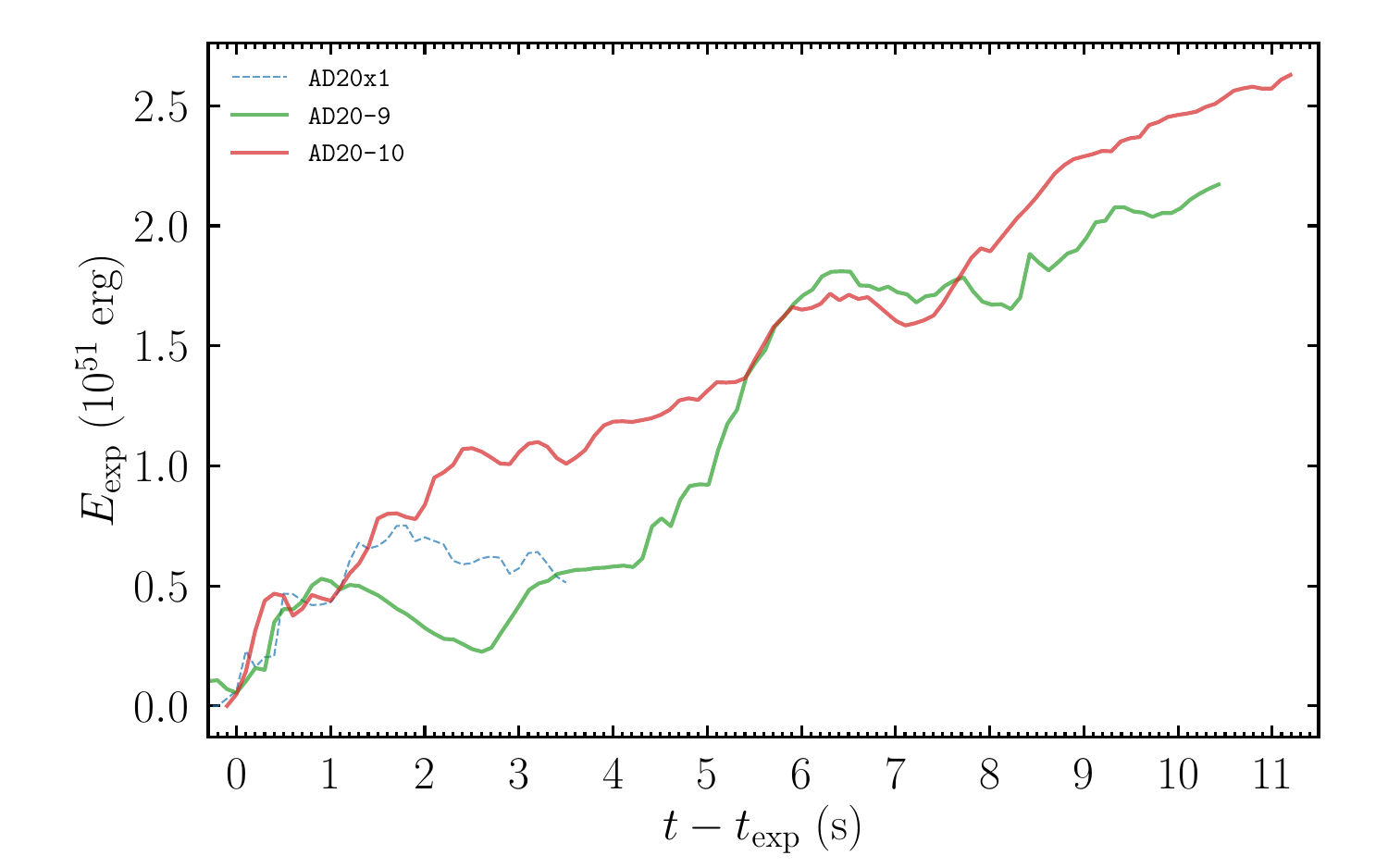}
\includegraphics[width=0.48\textwidth]{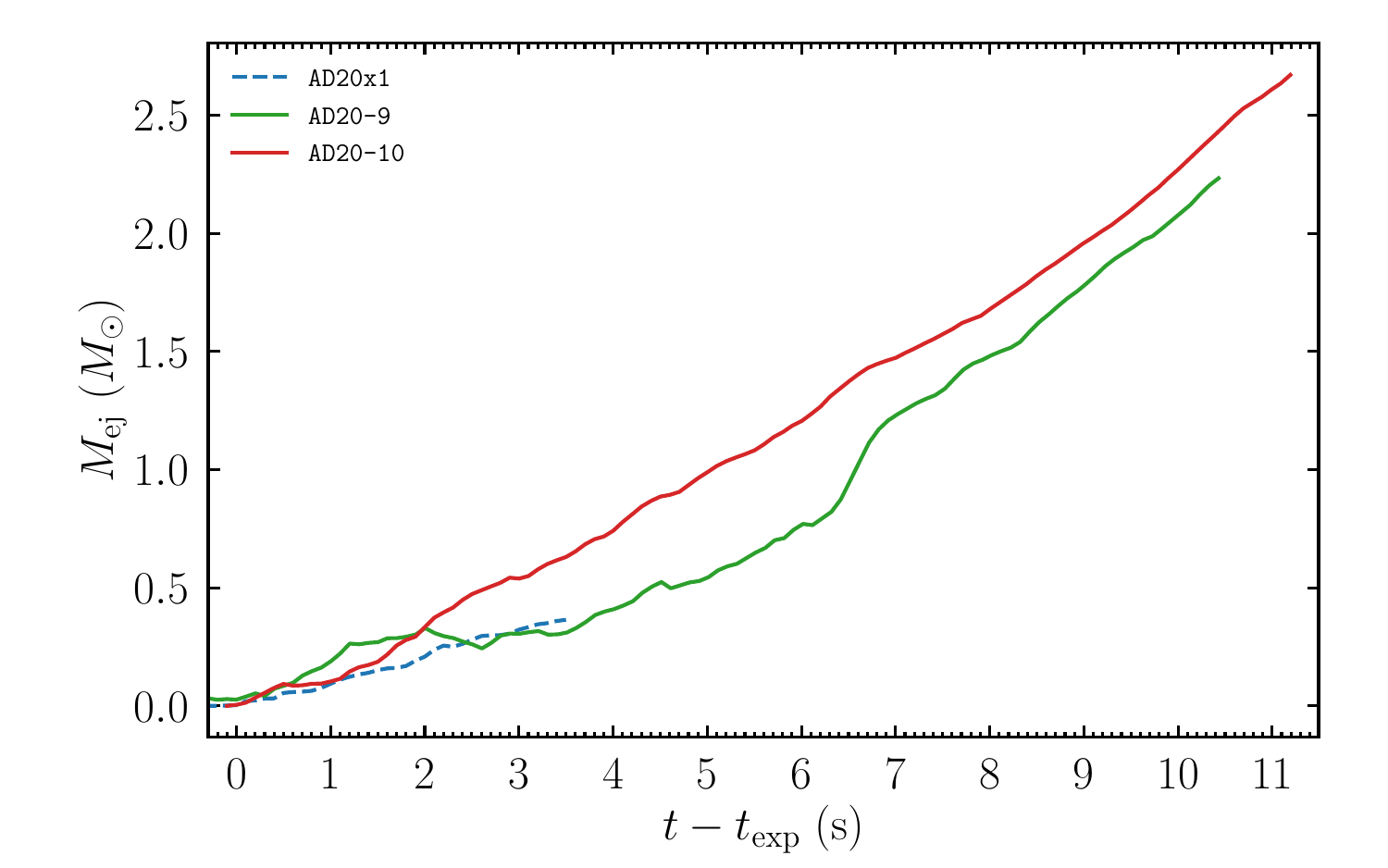}\\
\includegraphics[width=0.48\textwidth]{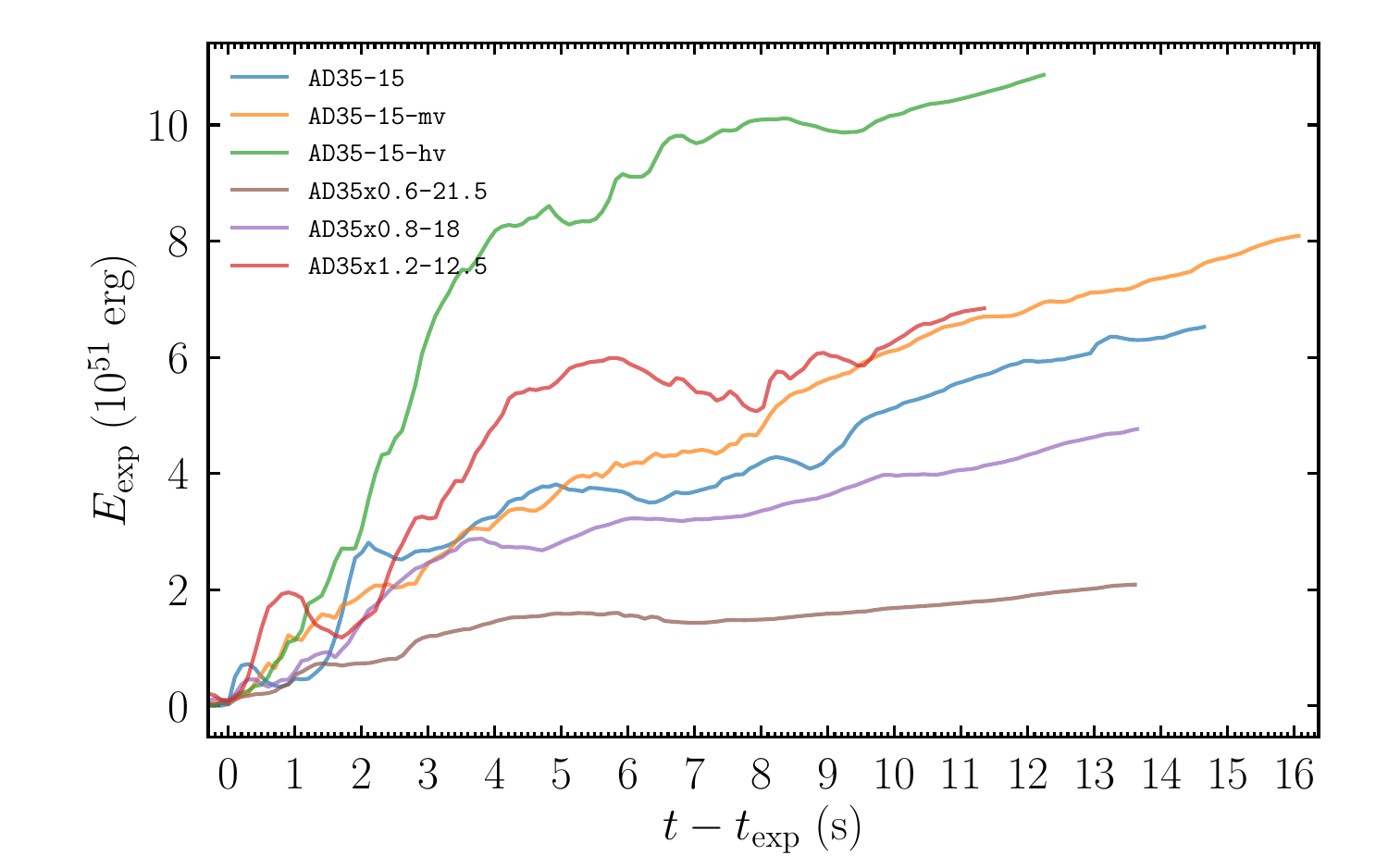}
\includegraphics[width=0.48\textwidth]{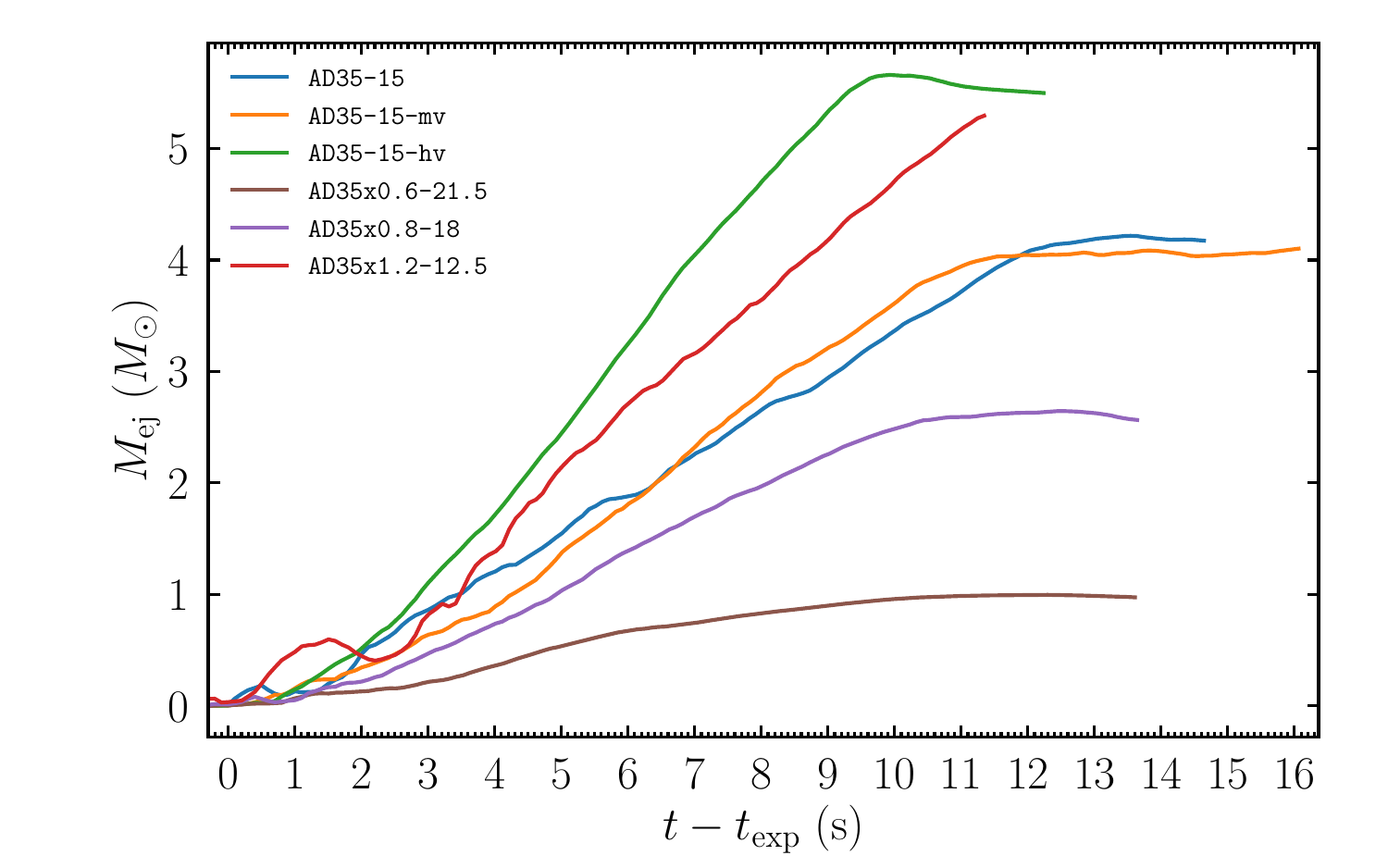}\\
\includegraphics[width=0.48\textwidth]{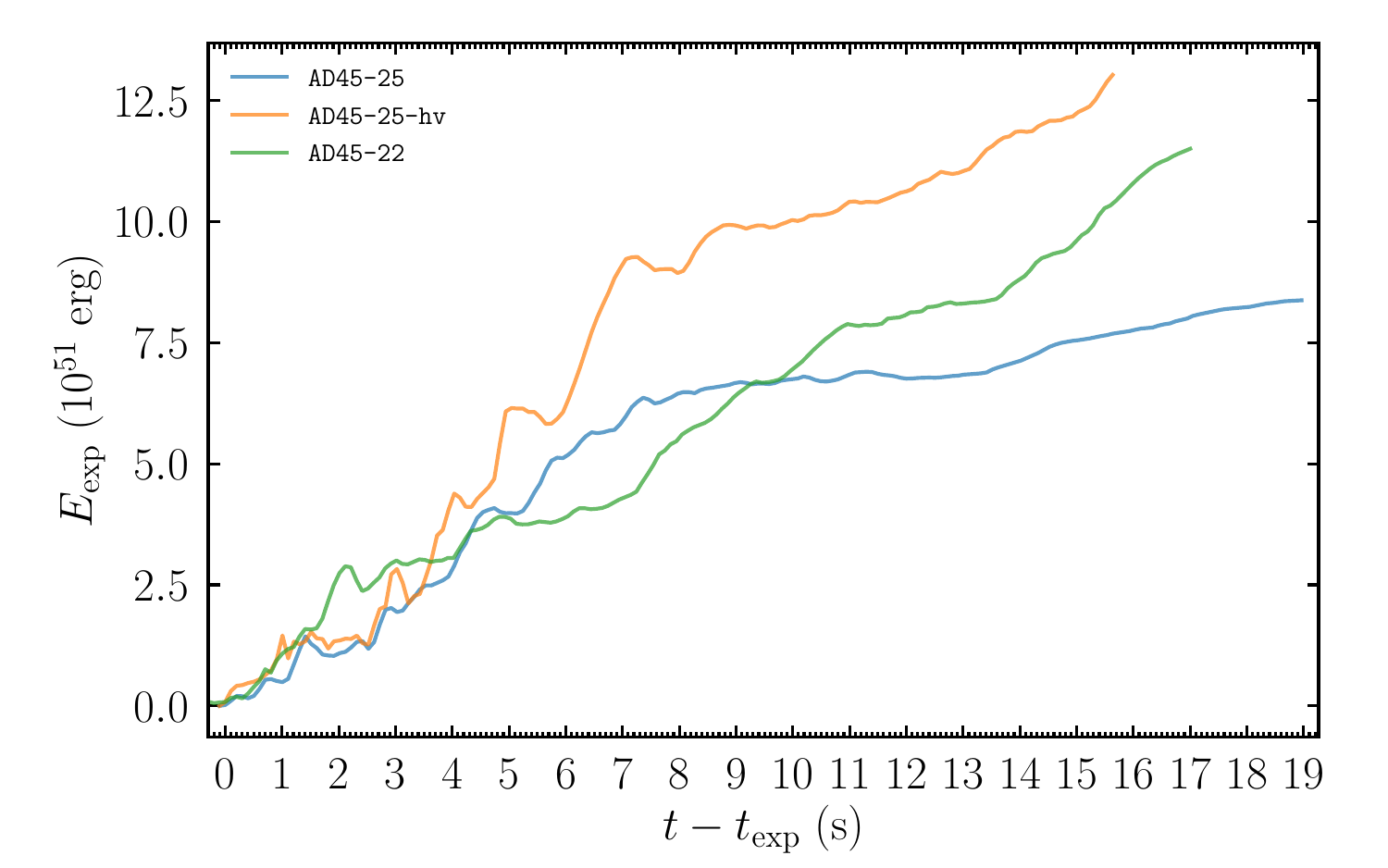}
\includegraphics[width=0.48\textwidth]{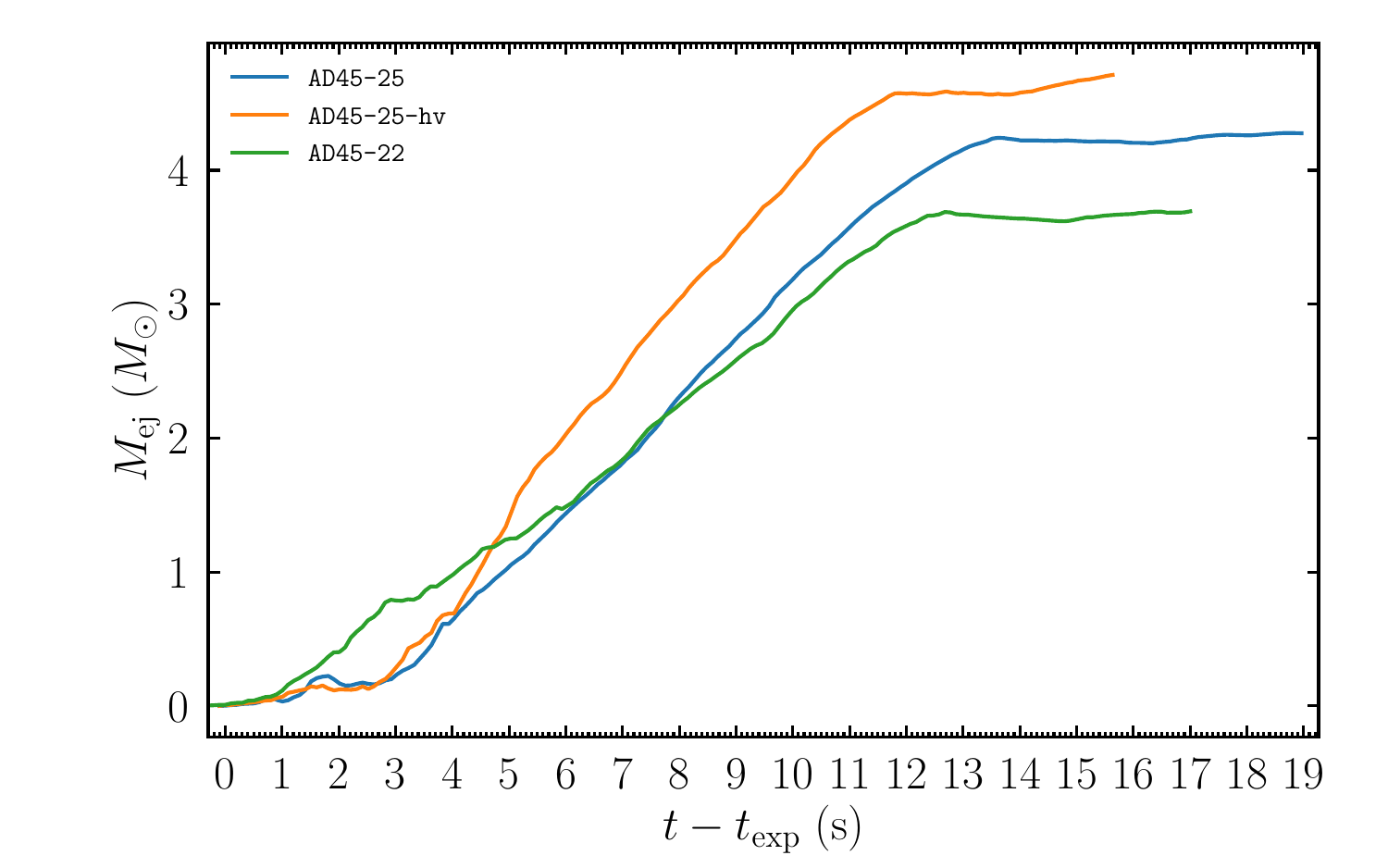}
\caption{Time evolution of the explosion energy (left) and ejecta mass (right) for models of $M_\mathrm{ZAMS}=20M_\odot$ (upper panels), $35M_\odot$ (middle panels), and  $45M_\odot$ (lower panels). For $M_\mathrm{ZAMS}=20M_\odot$, we also plot the result in Ref.~\cite{Fujibayashi2022dec} by the dashed curves. 
}
\label{fig:exp}
\end{figure*}

\begin{table*}[t]
    \centering
    \caption{Summary of the quantities associated with the explosion for the models for which the simulation is performed for sufficiently long time: Time at the onset of the explosion measured from the torus formation time, $t_\mathrm{exp}$ (the values in the parenthesis denote the simulation time), explosion energy, $E_\mathrm{exp}$, and ejecta mass,  $M_\mathrm{ej}$, measured at the termination of the simulation, the ejecta velocity defined by $v_\mathrm{ej}=\sqrt{2E_\mathrm{exp}/M_\mathrm{ej}}$, and synthesized $^{56}$Ni mass $M_\mathrm{Ni}$. In the last two columns, we also list the mass of an ejecta component with the temperature satisfying $T > 5\times 10^9$\,K during the ejection process and the average value of the entropy per baryon for the ejecta. For model \texttt{AD35x0.5-21.5}, we do not find explosion. For most of the models, the explosion energy was still increasing at the termination of the simulations, and thus, the values shown here are considered as the lower bound.
    }
    \begin{tabular}{lccccccc}
    \hline\hline
        Model & ~$t_\mathrm{exp}$\,(s)~  & $E_\mathrm{exp}$\,($10^{51}$\,erg) & ~~$M_\mathrm{ej}$\,($M_\odot$)~~ 
        & $v_\mathrm{ej}$ ($10^9\,{\rm cm/s})$
        & $M_\mathrm{Ni}$\,($M_\odot$) &
        $M_{>5\,\mathrm{GK}}$\,($M_\odot$) & $\langle s\rangle /k_\mathrm{B}$ \\
        \hline
        \texttt{AD20-9}        & 3.8    (3.8) & 2.2 & 2.2 & 1.0 & 0.24 & 0.44 & 17 \\
        \texttt{AD20-10}       & $<$0.1 (0.1) & 2.6 & 2.6 & 1.0 & 0.20 & 0.44 & 17\\
        \texttt{AD35-15}       & 2.8    (7.1) & 6.5  & 4.2 & 1.2 & 0.18 & 0.55 & 23\\
        \texttt{AD35-15-hi}    & 2.0    (6.3) & 7.0 & 5.0 & 1.2 & 0.24 & 0.72 & 28\\
        \texttt{AD35-15-mv}    & 0.8    (5.1) & 8.1  & 4.1 & 1.4 & 0.41 & 1.02 & 26\\
        \texttt{AD35-15-hv}    & 0.5    (4.8) & 10.1 & 5.5 & 1.4 & 0.15 & 0.69 & 39\\
        \texttt{AD35x0.5-21.5} & --- & --- & --- & --- & --- & --- & ---\\
        \texttt{AD35x0.6-21.5} & 0.7    (9.2) & 2.1 & 1.0 & 1.5 & 0.04 & 0.16 & 34 \\
        \texttt{AD35x0.8-18}   & 0.8    (7.2) & 4.4 & 2.6 & 1.7 & 0.15 & 0.52 & 32 \\
        \texttt{AD35x1.2-12.5} & 3.9    (7.4) & 6.8 & 5.3 & 1.1 & 0.38 & 0.90 & 23 \\
        \texttt{AD45-22}       & 0.6    (5.9) & 11.5 & 3.7   & 1.8 & 0.28 & 0.95 & 33 \\
        \texttt{AD45-25}       & $<$0.1 (0.1) & 8.4 & 4.3    & 1.4 & 0.46 & 1.15 & 27 \\
        \texttt{AD45-25-hv}    & $<$0.1 (0.1) & 13.0 & 4.7& 1.7& 0.25 & 0.87 & 43 \\
        \hline
    \end{tabular}\\
    \label{tab:model2}
\end{table*}
\subsection{Ejecta mass and explosion energy}\label{secIIIB}

Figure~\ref{fig:exp} shows the time evolution of the explosion energy (left panels) and ejecta mass (right panels) for all the models studied in this paper (see also Table~\ref{tab:model2}) except for model \texttt{AD35x0.5-21.5}, for which explosion is not found in the simulation time. At the termination of the simulations, the explosion energy is still increasing for most of the models, and hence, the values listed in Table~\ref{tab:model2} are considered to be the lower bound. However, broadly speaking, we may conclude that (i) for $M_\mathrm{ZAMS}=20M_\odot$, the explosion energy is a few times $10^{51}$\,erg, i.e., comparable to or slightly larger than that of the ordinary supernovae, while (ii) for $M_\mathrm{ZAMS}=35M_\odot$ and $45M_\odot$, it is $\sim 10^{52}$\,erg, i.e., about one order of magnitude larger than the ordinary supernovae, for the original progenitor models with no modification of the angular momentum profile. 

The large explosion energy of the massive progenitor models stems from their relatively large compactness. As we already mentioned in Sec.~\ref{secII}, for the pre-collapse models of Ref.~\cite{Aguilera-Dena2020oct}, the compactness of the progenitor star $C_*=GM_*/(c^2R_*)$ is larger for the more massive stellar models. Broadly speaking, the mass infall rate during the collapse is proportional to $M_*/t_\mathrm{ff}\propto C_*^{3/2}$, where $t_\mathrm{ff}=\sqrt{R_*^3/M_*}$ is the free-fall timescale. Thus, the mass-infall rate is higher for the larger-compactness progenitor models. The higher mass-infall rate enhances the viscous and shock heating rates around the inner region of the disk/torus, which result in the larger explosion energy for the more massive progenitor models.

For models with larger values of $\alpha_\nu$, the explosion energy and ejecta mass are naturally larger. Fundamentally, the viscous effect should come effectively from the magnetohydrodynamical turbulence and hydrodynamical shear in the present context. Thus, the explosion energy and ejecta mass can be accurately determined only by a magnetohydrodynamics simulation. However, the present study indicates that the dependence of these quantities on $\alpha_\nu$ is not very strong; even for the $10/3$ times larger value of $\alpha_\nu$, the explosion energy and ejecta mass increase within a factor of 2. In particular, the explosion energy and ejecta mass show similar values for $M_\mathrm{ZAMS}=35M_\odot$ with $\alpha_\nu=0.03$ and 0.06. Therefore it is reasonable to conclude that the explosion energy can reach $E_\mathrm{exp}\sim 10^{52}$\,erg with the ejecta mass of $M_\mathrm{eje}=4$--$5M_\odot$ for the present choice of the massive progenitor stars, if the turbulent state is excited and the resulting effective viscosity with $\alpha_\nu=ø(10^{-2})$ is generated around the inner region of the accretion disk/torus. 

The modification of the initial angular momentum profile for the progenitor stars of $M_\mathrm{ZAMS}=35M_\odot$ has an impact on the explosion energy and ejecta mass, in particular for the case that we reduce it by more than 40\%. The ejecta mass decreases monotonically with the decrease of the initial angular momentum because the total mass outside the black hole is initially smaller and the mass of the resulting disk/torus becomes smaller for the smaller initial angular momentum. The ejecta mass becomes $\sim 1M_\odot$ for the reduction of the angular momentum by $40\%$ (model \texttt{AD35x0.6-21.5}) and smaller than $0.4M_\odot$ (i.e., $< M_{*,0}-M_\mathrm{BH,f}$) by the $50\%$ reduction (model \texttt{AD35x0.5-21.5}). For model \texttt{AD35x0.6-21.5}, the explosion energy is $\sim 2\times 10^{51}$\,erg, which is comparable to that of ordinary supernovae. This suggests that a rapid rotation as well as the large compactness of the progenitor star is the key to the large explosion energy.

For the models of $M_\mathrm{BH}=20M_\odot$ and $45M_\odot$, we performed simulations with different initial black-hole mass. We find a fair agreement of the final values of explosion energy and ejecta mass, although their time evolution depends weakly on the initial setting. Thus, the ejecta-related quantities can be approximately obtained even if we start the simulations with black-hole mass larger than the value expected at the disk formation (see Sec.~\ref{secII}).



For $M_\mathrm{ZAMS}=20M_\odot$, we compare the present results with that in our previous paper~\cite{Fujibayashi2022dec}. We find that both the explosion energy and ejecta mass were underestimated in the previous study because the simulation time was too short. For obtaining the accurate explosion energy and ejecta mass for this case, we needed a long-term simulation with the duration of $\agt 10$\,s after the onset of the explosion. 


Even in the present study, the ejecta mass for $M_\mathrm{BH}=20M_\odot$ does not relax to a saturated value at the termination of the simulation. For this model, the expanding shock is still inside the computational domain, and a significant amount of unshocked, bound matter is present in the outer region of the star. The progenitor star for this model is less compact than the more massive progenitor stars, and hence, it takes more time (in units of $M_\mathrm{BH}$) to follow the ejecta generation. In the longer-term energy injection from the accretion torus, the ejecta mass may be increased to $M_{*,0}-M_\mathrm{BH,f} \sim 4M_\odot$. 

At the termination of the simulations for $M_\mathrm{ZAMS}=35M_\odot$ and $45_\odot$, we typically find $M_{*,0}-M_\mathrm{BH,f}-M_\mathrm{eje} \approx 1$--$2M_\odot$, which is still bound by the black hole. Since the black-hole mass increases slowly with time even at the termination of the simulations, most part of this mass will eventually fall into the black hole, and a fraction will be ejected from the system via the viscous heating and viscous angular momentum transport. However, this is a minor part compared with the matter ejected earlier. 

As mentioned in Sec.~\ref{secII}, we discard the stellar matter with $r > 10^5$\,km in our simulation for which the mass is $\sim 1M_\odot$. Thus the ejecta mass may be larger than those listed in Table~\ref{tab:model2} by this amount, but this possible increase is a small fraction of the numerical result of $M_\mathrm{ej}$ for most of the models.

\subsection{Nickel mass and predicted light curve}\label{secIIIC}

\begin{figure}
\includegraphics[width=0.49\textwidth]{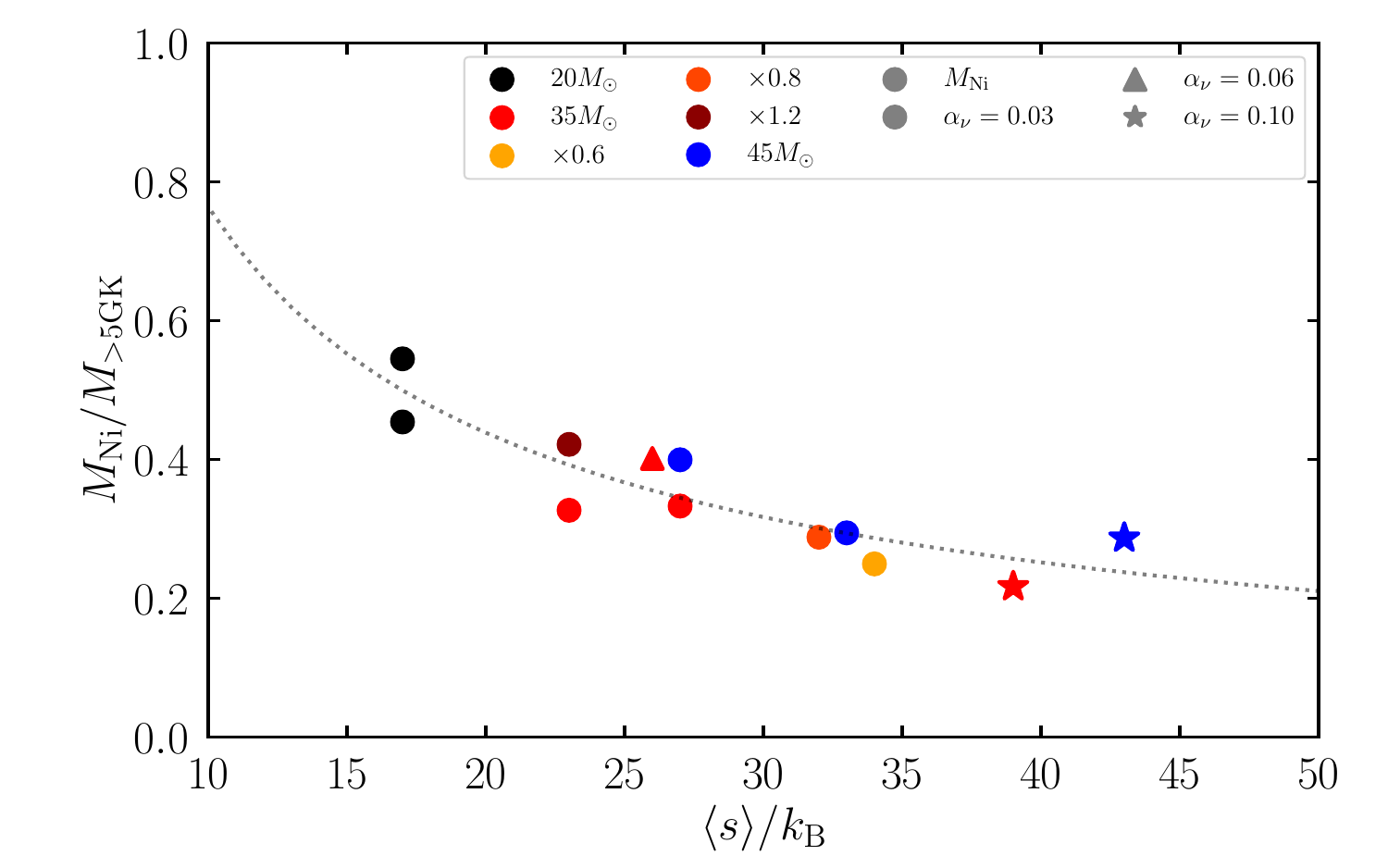}
\caption{$M_\mathrm{Ni}/M_{>5\,\mathrm{GK}}$ as a function of $\langle s\rangle/k_\mathrm{B}$. The dotted curve denotes $(\langle s \rangle/17k_\mathrm{B})^{-4/5}/2$.
}\label{fig:corr-Ni}
\end{figure}

\begin{figure*}
\includegraphics[width=0.49\textwidth]{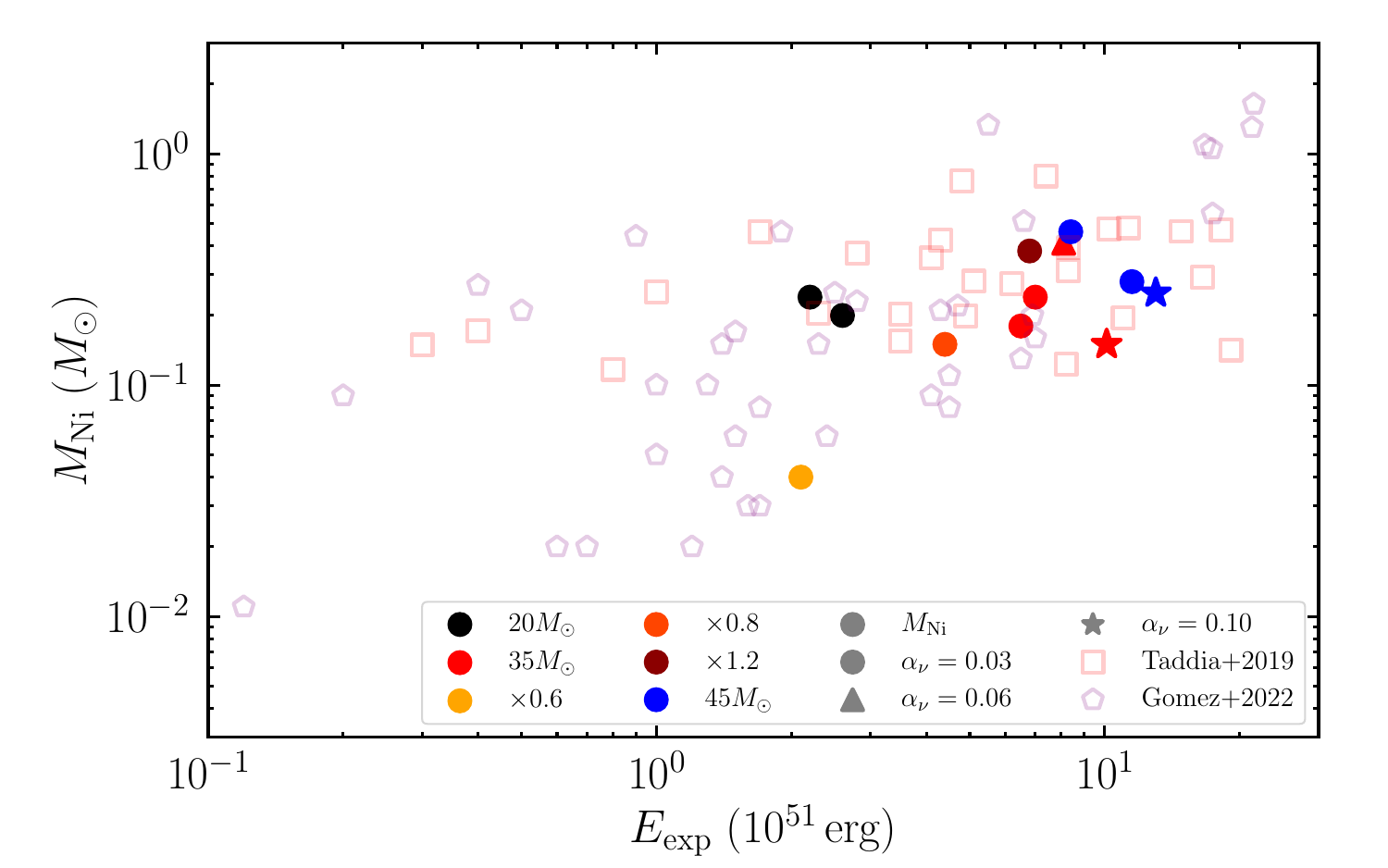}
\includegraphics[width=0.49\textwidth]{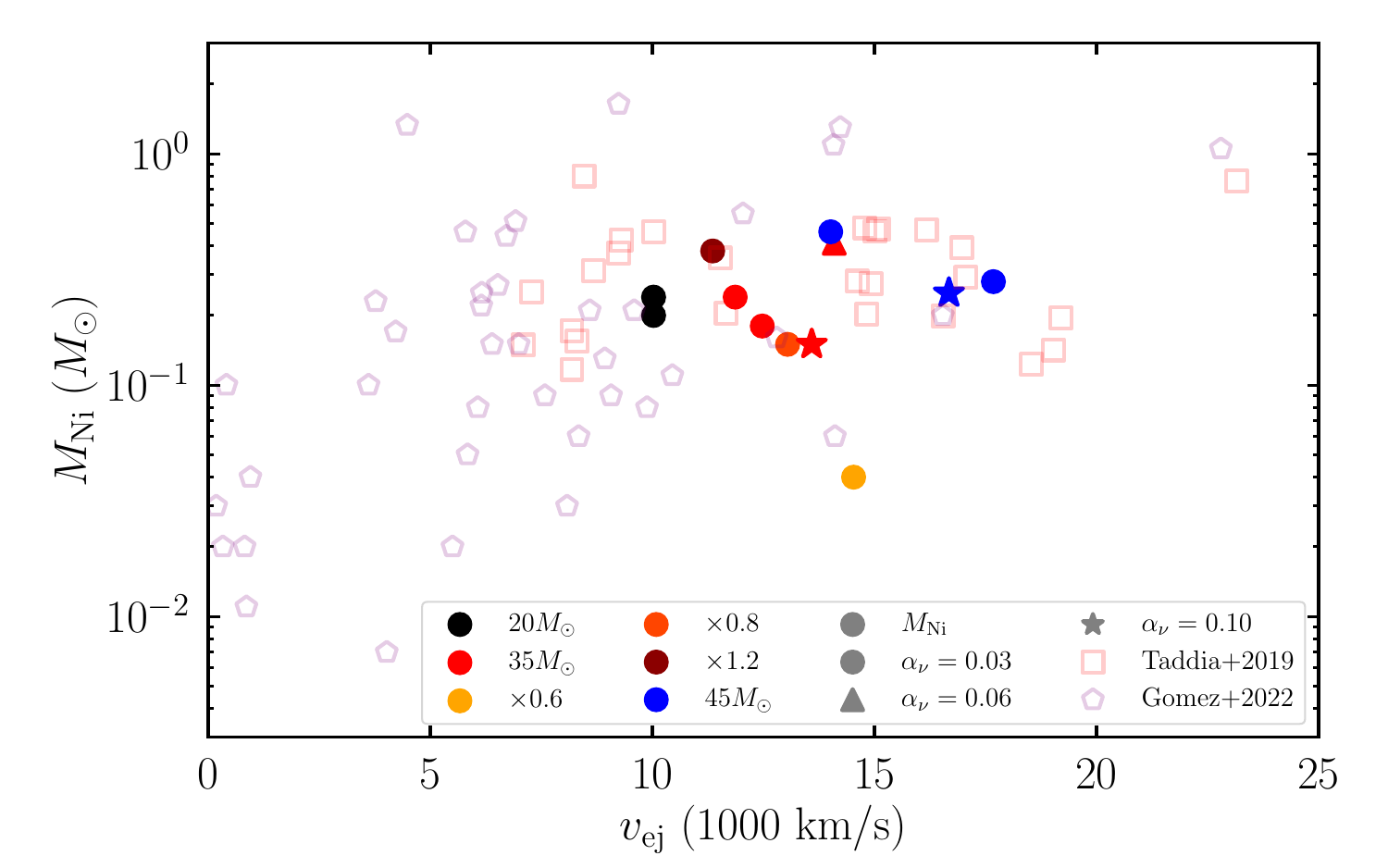}
\caption{
$M_\mathrm{Ni}$ as a function of the explosion energy $E_\mathrm{exp}$ (left) and average ejecta velocity $v_\mathrm{ej}$ (right). The open symbols denote the observational data for stripped-envelope supernovae, some of which are broad-lined type Ic supernovae, taken from Refs.~\cite{Taddia2019jan, Gomez2022dec}.
}\label{fig:Nimass}
\end{figure*}

Using the time evolution of the thermodynamical quantities on the tracer particles \cite{Fujibayashi2022dec}, post-process nucleosynthesis calculations are performed with a open-source nuclear reaction network code \texttt{torch} \cite{Timmes2000jul} with 495 isotopes, paying particular attention to the $^{56}$Ni production.

Table~\ref{tab:model2} lists the mass of $^{56}$Ni, $M_\mathrm{Ni}$, for selected models. The $^{56}$Ni mass is found to be always larger than $0.15M_\odot$ and $\sim 3$--$11\%$ of the total ejecta mass for all the models except for the models with significant angular momentum reduction (\texttt{AD35x0.5-21.5} and \texttt{AD35x0.6-21.5}). The $^{56}$Ni mass does not have strong correlation with the ejecta mass because the $^{56}$Ni production efficiency depends strongly on the thermal history of the matter during the explosion. In Table~\ref{tab:model2}, we also show the mass of the ejecta that experiences a state with $T > 5$\,GK ($=5\times10^{9}$\,K), $M_{>5\,\mathrm{GK}}$, and the average entropy per baryon, $\langle s\rangle/k_\mathrm{B}$, for the ejecta. The $^{56}$Ni production primarily occurs for $T\gtrsim5$~GK, while it is suppressed for the ejecta with a high entropy per baryon~\cite{Surman2011dec}. No clear correlation between $M_\mathrm{Ni}$ and the viscous coefficient is found (compare the results for models \texttt{AD35-15}, \texttt{AD35-15-mv}, and \texttt{AD35-15-hv}). This stems from the fact that the high viscous heating can enhance not only the fraction of the ejecta with $T>5$\,GK, but also the entropy per baryon. In our results, the $^{56}$Ni mass is approximately written as (see Fig.~\ref{fig:corr-Ni})
\begin{align}
M_\mathrm{Ni}\approx \frac{M_\mathrm{>5~GK}}{2}\bigg(\frac{\langle s \rangle}{17k_\mathrm{B}}\bigg)^{-4/5}.
\end{align}
It is also worth pointing out that $M_\mathrm{>5\,GK}$ is by more than a factor of $\sim 2$ larger than $M_\mathrm{Ni}$ for the models studied in this paper. Thus, $M_\mathrm{>5\,GK}$ overestimates the $^{56}$Ni mass for the present models.

By contrast, a clear correlation is found between $M_\mathrm{Ni}$ and the angular momentum of the progenitor stars for the $M_\mathrm{ZAMS}=35M_\odot$ model; larger angular momentum results in the larger $^{56}$Ni mass. This correlation stems from the larger mass and lower entropy per baryon of the ejecta for the larger initial angular momentum. The latter is associated with the difference in the evolution of the torus before the explosion sets in. For larger-angular-momentum models \texttt{AD35-15} and \texttt{AD35x1.2-12.5}, the explosion takes place after a quasi-stationary NDAF phase of the torus, during which neutrino emission extracts the entropy of the torus efficiently. In addition, the explosion after the quasi-stationary phase is less violent~\cite{Fujibayashi2022dec}. These factors result in the lower entropy of the ejecta. This situation is in clear contrast with those for smaller-angular-momentum models \texttt{AD35x0.6-21.5} and \texttt{AD35x0.8-18.0}, for which the explosion takes place in a relatively short timescale after the formation of the torus because of the lower neutrino cooling efficiency and lower ram pressure of infalling matter. For these models, a high entropy generated by the shock dissipation at the formation of the torus is directly reflected in that of the ejecta.

\begin{figure}
\includegraphics[width=0.49\textwidth]{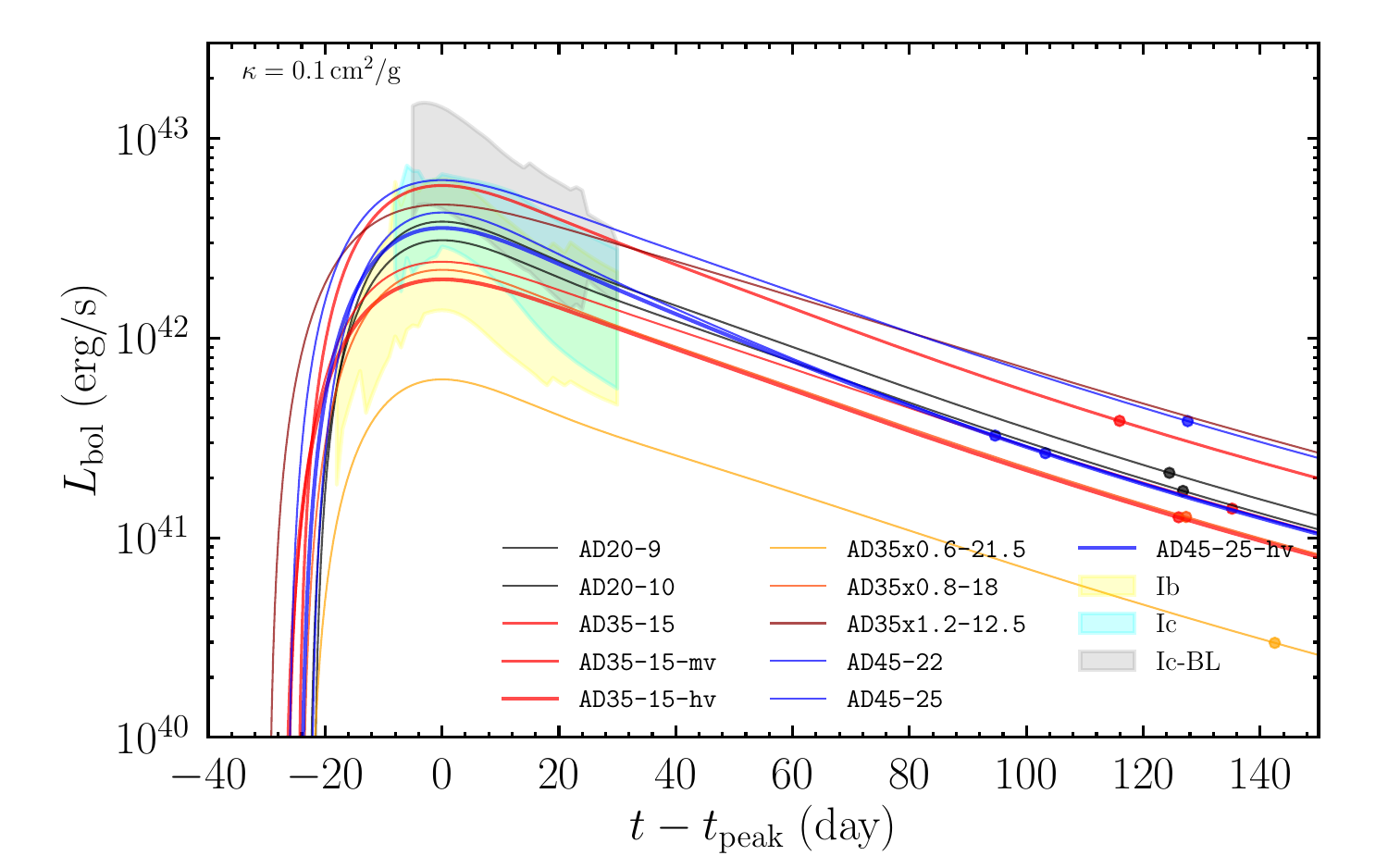}
\caption{
Bolometric light curves for all exploded models in this paper. Light curves for different models are plotted in different colors and line thicknesses. The filled circles along each curve indicate the time at which the ejecta becomes optically thin to thermal photons. The shaded regions denote templates of the bolometric light curves with standard deviations for type Ib, Ic, and Ic-BL taken from Ref.~\cite{Lyman2016mar}.
}\label{fig:Lbol}
\end{figure}

For the $M_\mathrm{ZAMS}=45M_\odot$ models, the $^{56}$Ni mass is larger, $\geq 0.25M_\odot$, reflecting the large mass fraction of the high-temperature ejecta component. The larger values of $M_{>\mathrm{5\,GK}}$ for these models result from the earlier explosion than for less massive progenitor models (see Sec.~\ref{secIIIA}). A significant difference is found between the results of models \texttt{AD45-22} and \texttt{AD45-25} in spite of the facts that for these models the explosion energy and ejecta mass show similar values. This illustrates that the $^{56}$Ni mass depends sensitively on the thermal condition of the ejecta.

Figure~\ref{fig:Nimass} displays the $^{56}$Ni mass as a function of the explosion energy (left panel) and the average ejecta velocity (right panel). Together with the numerical results shown by the filled symbol, we plot the observational data for stripped-envelope supernovae, some of which are broad-lined type Ic supernovae, taken from Refs.~\cite{Taddia2019jan,Gomez2022dec}, by the open symbols. It is found that our numerical results reproduce the relations between $M_\mathrm{Ni}$ and $E_\mathrm{exp}$ or $M_\mathrm{Ni}$ and $v_\mathrm{ej}$ for high-energy supernovae with $E_\mathrm{exp}=2$--$10 \times 10^{51}$\,erg and with $v_\mathrm{ej}=1$--$2\times 10^9$\,cm/s, suggesting that a fraction of these supernovae may be driven by the explosion from a torus surrounding a massive black hole of $M_\mathrm{BH}\approx 10$--$30M_\odot$.


Using the explosion energy, ejecta mass, and $^{56}$Ni mass as input parameters, we derive model light curves for the supernova-like explosion using the Arnett's model~\cite{Arnett1982}. In this modelling, we use the same prescription as described in our previous paper~\cite{Fujibayashi2022dec}. The resulting light curves are displayed in Fig~\ref{fig:Lbol}.
As predicted from the explosion energy, ejecta mass, and $^{56}$Ni mass, the peak luminosity and timescale of the luminosity decline for most of the models are in good agreement with the observed data for high-energy supernovae like the broad-lined type Ic supernovae or type Ib/Ic supernovae. For model \texttt{AD35x0.6-21.6}, the peak luminosity is lower than those for other models due to the smaller ejecta mass and explosion energy, indicating that a rapid rotation may be necessary to reproduce the brightness of high-energy supernovae. 

We note that the luminosity predicted by the Arnett model for given $^{56}$Ni mass may be underestimated by a factor of a few (see Refs.~\cite{Dessart2015oct,Dessart2016may,Khatami2019jun}). Thus, the explosion models presented in this paper may show more luminous light curves than in Fig.~\ref{fig:Lbol}, i.e., most of them may be good models for broad lined type Ic supernovae, as Fig.~\ref{fig:Nimass} indicates. To clarify this point, we need a more detailed radiation transfer study for deriving the light curves in follow-up work.

\section{Summary}\label{secV}

We studied the fate after the collapse of rotating massive stars that form a black hole and a disk/torus by performing a neutrino-radiation viscous-hydrodynamics simulation in general relativity and employing the stellar evolution models by Aguilera-Dena et al.~\cite{Aguilera-Dena2020oct} as initial data. Specifically, we employed rapidly rotating and compact progenitor stars as base models and constructed a system of a spinning black hole and infalling matter as the initial conditions. For most of the models we employed, a system of a black hole surrounded by a massive torus is formed during the time evolution. 

Due to the viscous heating as well as shock heating around the surface of the torus, thermal energy is generated and becomes the source for the explosion of the system. For the massive models ($M_\mathrm{ZAMS}=35M_\odot$ and $45M_\odot$), the ejecta mass is 4--$5 M_\odot$ and the explosion energy is $\sim 10^{52}$\,ergs, i.e., much larger than typical supernovae. The explosion energy is enhanced for larger viscous coefficients. By contrast, the explosion energy for the $20M_\odot$ model is of order $10^{51}$\,erg. The primary reason for this difference is that for the more massive models, the compactness of the progenitor stars is larger, the mass infall rate to the central part is higher, and as a result, the viscous and shock heating efficiency are enhanced to get large explosion energy. 

For $M_\mathrm{ZAMS}=35M_\odot$, we performed simulations artificially varying the initial angular momentum for a fairly wide range. For its change by $\pm 20\%$, the explosion energy and ejecta mass do not vary significantly. However, for the reduction by 50\%, we did not find the torus formation and explosion in our simulation time, although a small-mass disk is formed. This indicates that for high-energy explosion from the torus, a rapid rotation of the progenitor stars that results in a rapidly spinning black hole with $\chi \agt 0.7$ and a massive torus with mass $\agt 1M_\odot$ is necessary. 

For the simulations with the original progenitor models of Ref.~\cite{Aguilera-Dena2020oct}, the final black-hole spin is always 0.75--0.85, and thus, a rapidly spinning black hole is the outcome. The final black-hole mass is $\approx 10$--$30M_\odot$, which are 50--60\% of the progenitor mass. Even for the model with initially reduced angular momentum (model \texttt{AD35x0.5-21.5}) the final dimensionless spin is $\approx 0.6$. Since the black-hole dimensionless spin is high, in the presence of electromagnetic fields, the Blandford-Znajek effect is likely to play an important role~\cite{Blandford1977} for launching an energetic jet or outflow along the spin axis of the black hole. If a relativistic jet is produced, a gamma-ray burst will be also launched~(see Refs.~\cite{Komissarov:2004ms,Bromberg:2015wra,Gottlieb:2021srg} for simulation works). Our present explosion models may naturally explain the association between the gamma-ray burst and supernova-like explosion~\cite{Cano2017a} if a jet is really launched. To demonstrate that a relativistic jet is indeed launched, it is necessary to perform a magnetohydrodynamics simulation, which is one of our follow-up works to be done. In the presence of a jet, energy available for the explosion and $^{56}$Ni production is additionally injected, and also, observed relativistic motion in supernova-associated gamma-ray bursts will be naturally modelled~\cite{Cano2017a}. Exploring this additional effect is an important subject for developing a model for supernova-associated gamma-ray bursts.

For model \texttt{AD35x0.5-21.5}, energetic explosion from the torus is not found although a fairly rapidly spinning black hole is formed. In such a case, a gamma-ray burst may be launched in the presence of a strong magnetic field penetrating the black hole, while supernova-like explosion is likely absent. A wide variety of the final outcomes, which the present work illustrates, suggest that there may be a variety of possibilities on the high-energy phenomena depending on the initial angular momentum profiles in the progenitor stars. 

For the case that an explosion occurs, an appreciable amount of $^{56}$Ni is synthesized. We find that the $^{56}$Ni mass is always larger than $0.15M_\odot$ and $\sim 3$--$11\%$ of the total ejecta mass for rapidly rotating progenitor stars. For the models with reduced angular momentum, the $^{56}$Ni mass is significantly smaller. This illustrates that rapidly rotating progenitor stars are necessary for the significant $^{56}$Ni production. 

The relations between the explosion energy and $^{56}$Ni mass and between the average ejecta velocity and $^{56}$Ni mass are similar to the observational data for stripped-envelope supernovae with large explosion energy $>10^{51}$\,erg. As a natural consequence, the model light curves derived from our numerical results are also in good agreement with the observational data. This suggests a possibility that some of high-energy stripped-envelope supernovae may take place from a system of a spinning black hole and a massive torus. As discussed above, a gamma-ray burst is likely to accompany with such supernovae if a strong magnetic field penetrating the spinning black hole is developed. Therefore, supernova-associated gamma-ray bursts may be naturally explained in this model. 

\acknowledgements

We deeply thank Koh Takahashi for helpful discussions and David Aguilera-Dena for providing their stellar evolution models. We also thank Keiichi Maeda and Nozomu Tominaga for their helpful comments. Numerical computation was performed on Sakura, Momiji, Cobra, and Raven clusters at Max Planck Computing and Data Facility. This work was in part supported by Grant-in-Aid for Scientific Research (grant Nos.~20H00158 and 23H04900) of Japanese MEXT/JSPS.


\appendix

\section{Initial data for collapsing stars onto a spinning black hole} \label{A1}

We consider an axisymmetric initial data with the line element written in the form
\beqn
dl^2 = \psi^4 \hat\gamma_{ij}dx^i dx^j=
\psi^4 \left[e^{2q}(dR^2 +dz^2) + R^2d\varphi^2\right],\nonumber \\
\label{eq1}
\eeqn
where $\hat \gamma_{ij}$ is the conformal three metric and $\psi$ is a conformal factor, both of which are functions of $R$ and $z$.  We suppose that $q$ is a given function of $R$ and $z$. We require that the metric reduces to that of Kerr black holes in the quasi-isotropic coordinates in the absence of matter~\cite{Krivan1998nov}, i.e., 
\beqn
\psi &=&\psi_{\rm K}={\Xi_{\rm K}^{1/4} \over r^{1/2}\Sigma_{\rm K}^{1/4}},\\
e^q &=& e^{q_{\rm K}}={\Sigma_{\rm K} \over \Xi_{\rm K}^{1/2}},
\eeqn
where
\beqn
\Xi_{\rm K}&=&(r_{\rm K}^2 +a^2)\Sigma_{\rm K} + 2Ma^2 r_{\rm K}\sin^2\theta,\\
\Sigma_{\rm K}&=&r_{\rm K}^2 + a^2\cos^2\theta,
\eeqn
$M$ is the black-hole mass, $a$ is the black-hole spin, $r_{\rm K}$ is the radial coordinate in the Boyer-Lindquiest coordinates of Kerr black holes, $r=\sqrt{R^2 + z^2}$, and $\tan\theta=R/z$.  The relation between $r_{\rm K}$ and $r$ is
\beq
r_{\rm K}=r + M + {r_{\rm s}^2 \over r},
\eeq
where $r_{\rm s}:=\sqrt{M^2-a^2}/2$ denotes the location of the black-hole horizon in the quasi-isotropic coordinates. In the following, we assume $q=q_{\rm K}$. We note that for $r\rightarrow 0$, $\Psi_{\rm K} \rightarrow r_{\rm s}/r$ and $q_{\rm K}\rightarrow 0$. 

From the extrinsic curvature $K_{ij}$, we define $\hat K_{ij}=\psi^2 K_{ij}$, $\hat K^i_{~j}=\psi^6 K^i_{~j}$, $\hat K^{ij}=\psi^{10} K^{ij}$, and the subscripts of $\hat K_{ij}$ is raised by $\hat \gamma^{ij}$. In the following, we assume that the trace of the extrinsic curvature is zero, i.e., $(\hat K^{RR} + \hat K^{zz})e^{2q} + \hat K^{\varphi\varphi}R^2=0$. Then, for the metric of Eq.~(\ref{eq1}), the momentum constraint is written in the form:
\beqn
&&    {1 \over R} \pa_R (R \hat K_{RR})+\pa_z \hat K_{Rz}
-(\hat K_{RR} + \hat K_{zz}) (\pa_R q -R^{-1})
\nonumber\\
&& ~~~~~~~~~~~~~~~~~~~~~~~~~~~~~~~~=8\pi J_R \psi^6 e^{2q},\\
&&    {1 \over R} \pa_R (R \hat K_{Rz})+\pa_z \hat K_{zz}
    -(\hat K_{RR} + \hat K_{zz}) \pa_z q 
\nonumber \\
&& ~~~~~~~~~~~~~~~~~~~~~~~~~~~~~~~~=8\pi J_R \psi^6 e^{2q},\\
&&    {1 \over R} \pa_R (R \hat K_{R\varphi})+\pa_z \hat K_{z\varphi}
    =8\pi J_\varphi \psi^6 e^{2q},
\eeqn
where $J_i=\alpha T^t_{~i}$ with $\alpha$ the lapse function and $T^{\mu\nu}$ the energy-momentum tensor. In the formalism presented here, we will give $J_i$ to determine the geometric quantities, and hence, we do not have to specify $\alpha$. 

We then write the conformal-tracefree extrinsic curvature as
\beqn
\hat K_{ij}=\hat D_i W_j + \hat D_j W_i -{2 \over 3}\hat\gamma_{ij}
\hat D_k W^k + \hat K^{\rm K}_{ij},
\eeqn
where $\hat D_i$ is the covariant derivative with respect to $\hat\gamma_{ij}$, $W^i$ is a conformal three vector, i.e., $W_j=\hat\gamma_{jk}W^k$, and $\hat K^{\rm K}_{ij}$ is the contribution from the black hole, which is trancefree. Each component of $\hat K_{ij}$, necessary for the momentum constraint, is written as
\beqn
&&\hat K_{RR}=\pa_R W_R -{W_R \over R}-\pa_z W_z -2 W_R \pa_R q +
2W_z \pa_z q \nonumber \\
&&~~~~~~~+{1 \over 3} {\rm div} W,\nonumber \\
&&\hat K_{Rz}=\pa_R W_z + \pa_z W_R -2 W_R \pa_z q -
2W_z \pa_R q,\nonumber \\
&&\hat K_{zz}=\pa_z W_z -\pa_R W_R-{W_R \over R} + 2 W_R \pa_R q -
2W_z \pa_z q \nonumber \\
&&~~~~~~~+{1 \over 3} {\rm div} W,\nonumber \\
&&\hat K_{R\varphi}=\pa_R W_\varphi -2 {W_\varphi \over R}+\hat K^{\rm K}_{R\varphi},
\nonumber \\
&&\hat K_{z\varphi}=\pa_z W_\varphi+\hat K^{\rm K}_{z\varphi},
\eeqn
where ${\rm div} W=\pa_R W_R + W_R/R + \pa_z W_z$, 
\beqn
&& \hat K^{\rm K}_{R\varphi}={H_E R^3 \over r^5}+{H_F Rz \over r^4},\\
&& \hat K^{\rm K}_{z\varphi}={H_E R^2z \over r^5}-{H_F R^2 \over r^4},
\eeqn
and $H_E$ and $H_F$ are~\cite{Brandt1995jul1,Brandt1995jul2}
\beqn 
&& H_E={M a \left[ (r_{\rm K}^2 -a^2) \Sigma_{\rm K} + 2 r_{\rm K}^2
    (r_{\rm K}^2 +a^2) \right] \over \Sigma_{\rm K}^2},\\
&& H_F=-{2 M a^3 r_{\rm K}
  \sqrt{r_{\rm K}^2 -2 Mr_{\rm K} +a^2} \sin^2\theta\cos\theta
  \over \Sigma_{\rm K}^2}.~~~~~~~
\eeqn
Here, $\hat K^{\rm K}_{ij}$ satisfies the $\varphi$-component of the momentum constraint for $J_\varphi=0$
\beqn
    {1 \over R} \pa_R (R \hat K_{R\varphi}^{\rm K})
    +\pa_z \hat K_{z\varphi}^{\rm K}=0. 
\eeqn    

Then the equations for $W_i$ are written as
\beqn
&& \left[ \Delta -{1 \over R^2} \right] W_R
+{1 \over 3}\pa_R ({\rm div} W) \nonumber \\
&&~~~-2\left( \pa^2_Rq + \pa^2_zq\right) W_R
-\left({8 \over 3} {\rm div}W -{2W_R \over R}\right)\pa_R q
\nonumber \\
&&~~~+2 \left(\pa_R W_z + {W_z \over R} -\pa_z W_R \right)\pa_z q
\nonumber \\
&&~~~=8\pi J_R \psi^6 e^{2q},\\
&& \Delta W_z +{1 \over 3}\pa_z ({\rm div} W) \nonumber \\
&&~~~-2\left( \pa^2_Rq + \pa^2_zq\right) W_z
-\left({8 \over 3} {\rm div}W -{2W_R \over R}\right)\pa_z q
\nonumber \\
&&~~~-2 \left(\pa_R W_z + {W_z \over R} -\pa_z W_R \right)\pa_R q
\nonumber \\
&&~~~=8\pi J_z \psi^6 e^{2q},\\
&& \left[ \Delta -{1 \over R^2} \right] W^{\bar{\varphi}}
=8\pi J_\varphi \psi^6 e^{2q} R^{-1}, \label{eqWphi}
\eeqn
where $W^{\bar{\varphi}}:= W^\varphi/R$ and $\Delta$ denotes the flat Laplacian, 
\beqn
\Delta = \pa^2_R + {1 \over R}\pa_R + \pa^2_z. 
\eeqn
For a given function of $J_\varphi \psi^6 e^{-2q}$, the equation for $W^{\bar{\varphi}}$ is solved with the outer boundary condition of $W^{\bar{\varphi}} \propto r^{-2}$ and the inner boundary conditions, $W^{\bar{\varphi}}\propto R$ for $R\rightarrow 0$ and $\pa_z W^{\bar{\varphi}}=0$ at $z=0$. 

To simplify the procedure for the numerical solution of $W_R$ and $W_z$, we may rewrite these variables using (see, e.g., Ref.~\cite{Shibata2016a} for a similar formulation in Cartesian coordinates)
\beqn
W_i=B_i - {1 \over 8} \pa_i (\chi + B_R R + B_z z),
\eeqn
where $\chi$ and $B_i$ are new functions to be solved instead of $W_R$ and $W_z$, and $i$ denotes $R$ or $z$. With this prescription, we find
\beqn
&& \left[\Delta -{1 \over R^2} \right] W_R +{1 \over 3}\pa_R ({\rm div} W)
\nonumber \\
&&= \left[\Delta  -{1 \over R^2} \right] B_R \nonumber \\
&&~-{1 \over 6}\pa_R \left[\Delta \chi + R (\Delta -R^{-2}) B_R+ z \Delta B_z\right]
\eeqn
and 
\beqn
&& \Delta W_z +{1 \over 3}\pa_z ({\rm div} W) \nonumber \\
&&= \Delta B_z -{1 \over 6}\pa_z \left[\Delta \chi
  + R (\Delta -R^{-2}) B_R+ z \Delta B_z\right].~~~~~
\eeqn
Thus, by choosing the equation for $\Delta \chi$ as
\beqn
\Delta \chi = - R (\Delta -R^{-2}) B_R - z \Delta B_z,
\eeqn
we obtain the equations for $B_R$, $B_z$, and $\chi$ in simple forms as
\beqn
\left[ \Delta -{1 \over R^2} \right] B_R &=&S_R, \label{eqBr}\\
\Delta B_z &=&S_z, \label{eqBz}\\
\Delta \chi &=&-R S_R - zS_z, \label{eqchi}
\eeqn
where
\beqn
S_R&=&
2\left( \pa^2_Rq + \pa^2_zq\right) W_R
+\left(2 {\rm div}B -{2W_R \over R}\right)\pa_R q
\nonumber \\
&&-2 \left(\pa_R B_z + {W_z \over R} -\pa_z B_R \right)\pa_z q
\nonumber \\
&&+8\pi J_R \psi^6 e^{2q},\\
S_z&=&
2\left( \pa^2_Rq + \pa^2_zq\right) W_z
+\left(2 {\rm div}B -{2W_R \over R}\right)\pa_z q
\nonumber \\
&&+2 \left(\pa_R B_z + {W_z \over R} -\pa_z B_R \right)\pa_R q
\nonumber \\
&&+8\pi J_z \psi^6 e^{2q},
\eeqn
and
\beq
    {\rm div}B \left(={4 \over 3}{\rm div}W\right)
    =\pa_R B_R + {1 \over R} B_R + \pa_z B_z. 
\eeq
We note that in $S_R$ and $S_z$ the second spatial derivative of $B_R$, $B_z$, and $\chi$ is not present.

Because $S_R$ and $S_z$ fall off sufficiently rapidly in the far region (with $O(r^{-6})$), the elliptic equations (\ref{eqBr})--(\ref{eqchi}) can be solved in a straightforward manner with the outer boundary conditions
\beqn
B_R \propto {R \over r^3},~~~~~
B_z \propto {z \over r^3},~~~~~
\chi \propto {1 \over r}. 
\eeqn
The boundary conditions at $R=0$ are
\beqn
B_R \propto R,~~~~~\pa_R B_z=0=\pa_R \chi,
\eeqn
and the boundary conditions at $z=0$ are
\beqn
\pa_z B_R =0=\pa_z \chi,~~~~~B_z \propto z.
\eeqn
For the equation of $B_R$, it may be better to solve the equation for $B_{\bar R}=B_R/R$ to guarantee the boundary condition, $\pa_R B_{\bar R}=0$, at $R=0$. 
For this case the kernel operator of the equation becomes
\beq
\left(\pa^2_R + {3 \over R}\pa_R +\pa^2_z \right)B_{\bar R}={S_R \over R}. 
\eeq
Here, we note that $J_R \propto R$ and $q \propto \sin^2\theta$ at $\theta \rightarrow 0$, and thus, the regularity of $S_R/R$ at $R=0$ is guaranteed.

If we consider that $J_i \psi^6 e^{2q}$ is a given function, the Hamiltonian constraint is  solved for an obtained numerical solution of $\hat K_{ij}$. In this context, the Hamiltonian constraint is written as
\beqn
\Delta \psi ={1 \over 8} \psi e^{2q} \hat R -2\pi \rho_{\rm H} \psi^5 e^{2q}
-{1 \over 8\psi^7} \hat K_{ij}\hat K^{ij},\label{eqH}
\eeqn
where $\rho_{\rm H}=\alpha^2 T^{tt}$ and $\hat R$ is the Ricci scalar with respect to the given conformal metric, $\hat \gamma_{ij}$, i.e., $q=q_{\rm K}$. In the present context (e.g., Ref.~\cite{Shibata2007sep}), 
\beq
\hat R=-2 e^{-2q} (\pa^2_R +\pa^2_z)q. 
\eeq
We also note that we will consider to give $\rho_{\rm H}$ (not $T^{tt}$), and hence, we do not have to specify $\alpha$.  

For the decomposition of $\psi=\psi_{\rm K} + \phi$, Eq.~(\ref{eqH}) is rewritten as
\beqn
\Delta \phi &=&{1 \over 8} \phi e^{2q} \hat R -2\pi \rho_{\rm H} \psi^5 e^{2q}
\nonumber \\
&&-{1 \over 8\psi^7} \hat K_{ij}\hat K^{ij}
+{1 \over 8\psi_{\rm K}^7} \hat K^{\rm K}_{ij}\hat K^{{\rm K}ij},
\label{eqphi}
\eeqn
where we used
\beqn
\Delta \psi_{\rm K} ={1 \over 8} \psi_{\rm K} e^{2q} \hat R 
-{1 \over 8\psi_{\rm K}^7} \hat K^{\rm K}_{ij}\hat K^{{\rm K}ij}.
\eeqn
The boundary conditions for $\phi$ are
\beqn
&&\pa_r [r(\phi-1)]=0~~{\rm at}~~r\rightarrow \infty,\\
&&\pa_R \phi=0~~{\rm at}~~R=0,\\
&&\pa_z \phi=0~~{\rm at}~~z=0. 
\eeqn
For $r\rightarrow 0$, $\psi_{\rm K}\propto r^{-1}$, $K_{ij}^{\rm K}K^{{\rm K}ij} \propto r^{-6}$, and $\hat R \rightarrow 2a^2/r_{\rm s}^4$, the right-hand side of Eq.~(\ref{eqphi}) is regular anywhere.  Thus, it is also straightforward to solve this equation under the boundary conditions shown above.

For the perfect fluid,
\beq
T^{\mu\nu}=\rho h u^\mu u^\nu + P g^{\mu\nu},
\eeq
where $\rho$, $h$, $u^\mu$, $P$, and $g^{\mu\nu}$ are the rest-mass density, specific enthalpy, four velocity, pressure, and spacetime metric. Then we obtain
\beqn
\hat J_i:=J_i \psi^6 e^{2q} &=& \rho h \alpha u^t u_i \psi^6 e^{2q}=\rho_* h u_i,\\
S_0:=\rho_{\rm H}\psi^6 e^{2q}&=&\rho_* h (\alpha u^t)-P\psi^6 e^{2q},
\eeqn
where $\rho_*=\rho \alpha u^t \psi^6 e^{2q}$ is the weighted rest-mass density which satisfies the continuity equation,
\beq
\pa_t \rho_* + {1 \over R} \pa_R  \left(R\rho_* v^R\right)
+ \pa_z (\rho_* v^z)=0, 
\eeq
with $v^i=u^i/u^t$ and $\alpha u^t=\sqrt{1+\psi^{-4}\hat\gamma^{ij}u_iu_j}$. Thus, the total rest mass of the system is obtained by
\beq
M_* = 2\pi \int RdR dz \,\rho_*. 
\eeq
The angular momentum of the matter is also obtained by
\beq
J = 2\pi \int RdR dz \,\hat J_\varphi. 
\eeq
In numerical computation, $\left(\rho_*, Y_e, T, \hat{J}_\phi, u_R, u_z \right)$ are provided using the data of the collapsing matter (see Sec.~\ref{secII}), and the field equations, e.g.,  (\ref{eqBr}), (\ref{eqBz}), (\ref{eqchi}), and (\ref{eqphi}), are solved iteratively until the rest-mass density $\rho$ and all metric variables converge.

\section{Accuracy of the black-hole quantities}\label{A2}

\begin{figure}[t]
\includegraphics[width=0.49\textwidth]{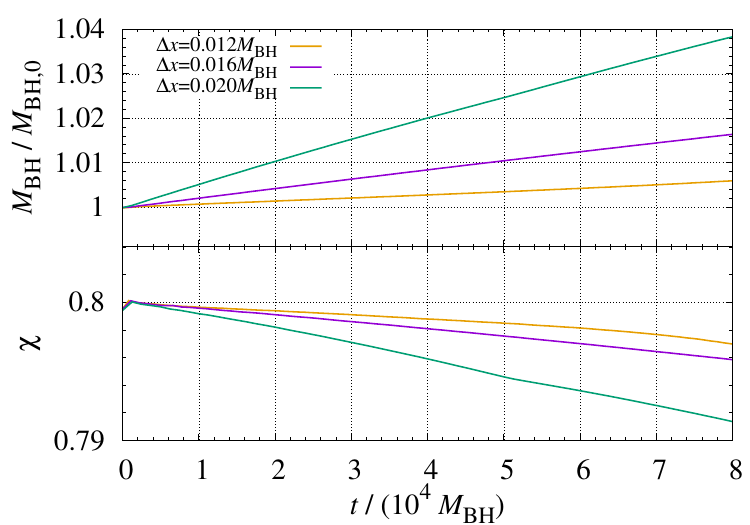}
\caption{Evolution of the mass (upper panel) and dimensionless spin (lower panel) of spinning black holes for $\chi=0.8$ with the grid resolutions of $\Delta x/M_\mathrm{BH}=0.012$, $0.016$, and $0.020$. 
}
\label{figapp}
\end{figure}

\begin{figure*}[t]
\includegraphics[width=0.49\textwidth]{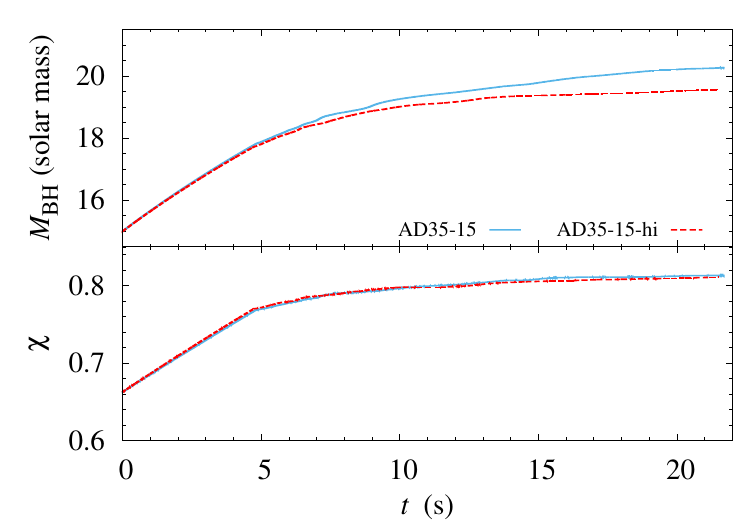}
\includegraphics[width=0.49\textwidth]{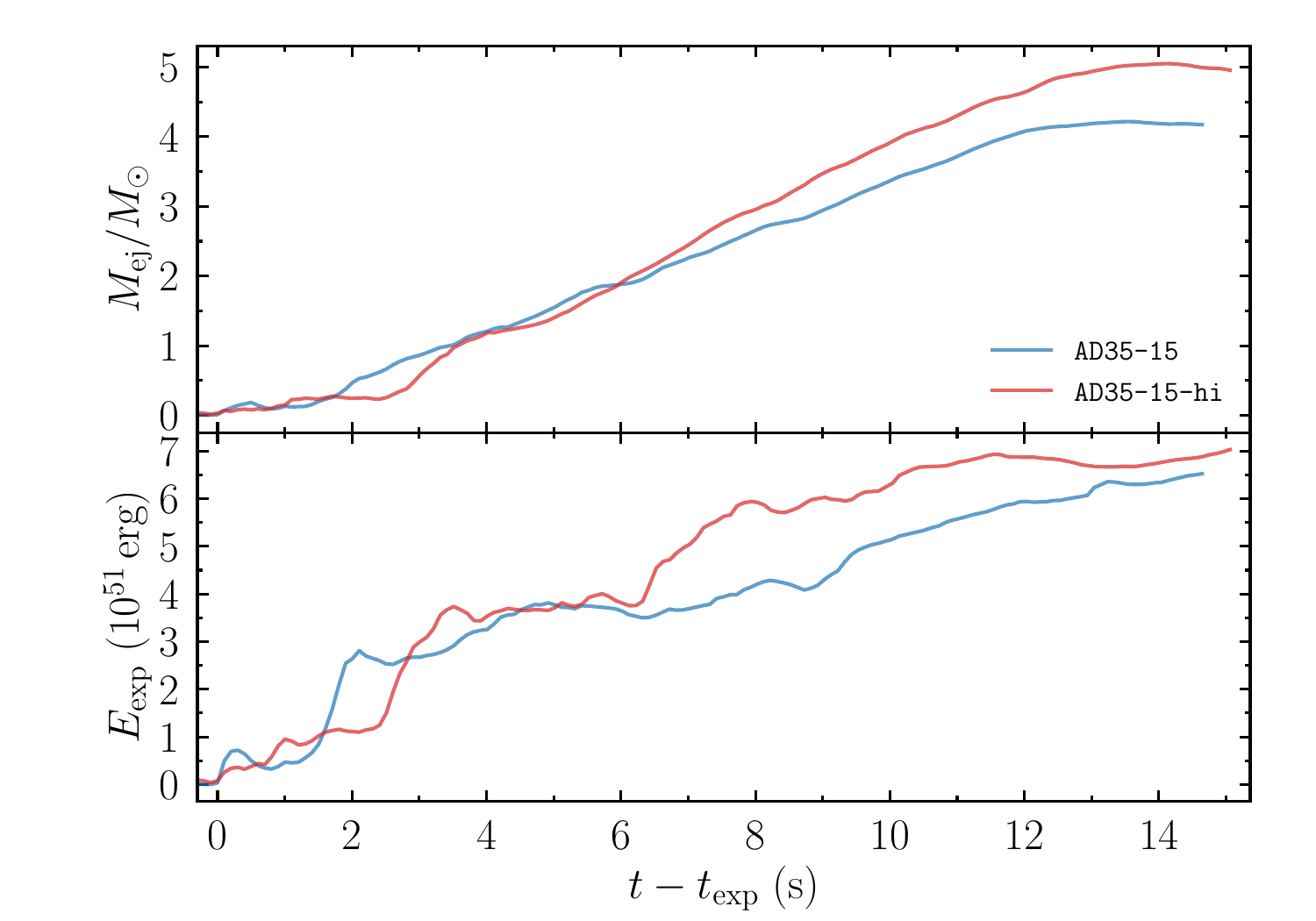}
\caption{Left: The same as the middle panel of Fig.~\ref{fig:BH} but for the comparison between the results of models \texttt{AD35-15} (solid curves) and \texttt{AD35-15-hi} (dashed curves). 
Right: The ejecta mass (upper panel) and explosion energy (lower panel) for models \texttt{AD35-15} and  \texttt{AD35-15-hi}. 
}
\label{figapp2}
\end{figure*}

To ascertain numerical accuracy in evaluating the mass and dimensionless spin of black holes, we evolve isolated spinning black holes using similar grid resolutions to those used in the present work, initially preparing a Kerr black hole in quasi-isotropic coordinates~\cite{Krivan1998nov} with $\chi=0.8$. Numerical evolution is carried out until $t=80,000M_\mathrm{BH}$. To save the computational costs, the outer boundary is located at $\approx 800M_\mathrm{BH}$ along each axis. The simulations are performed for $\Delta x/M_\mathrm{BH}=0.012$, $0.016$, and $0.020$ which are employed for the uniform grid zone with $x\leq 0.72M_\mathrm{BH}$ where $x$ denotes $R$ or $z$. For $x > 0.72M_\mathrm{BH}$ the grid spacing is increased with the rate of 1.01 as in viscous hydrodynamics simulations. In this section, the results are shown in units of $M_\mathrm{BH}=1$ (with $c=1=G$). For example, for $M_\mathrm{BH}=15M_\odot$, $80,000M_\mathrm{BH}\approx 5.9$\,s and $800M_\mathrm{BH} \approx 1.8\times 10^4$\,km. 

Figure~\ref{figapp} shows the evolution of the mass and dimensionless spin. A bump found at $t\approx 1,600M_\mathrm{BH}$ is due to a slight reflection of numerical errors from the outer boundary: In this test simulations, the initial data are Kerr black holes in the quasi-isotropic coordinates, and thus, during the time evolution, the metric form is varied due to the change of the slicing, approaching those on the limiting hypersurface (trumpet hypersurface). During this variation, the gauge modes are propagated outward with the speed of light and some of the modes are reflected at the outer boundary toward the inner region causing a high-frequency numerical noise. This oscillation spuriously and slightly perturbs the horizon in particular for the high-resolution runs, but the oscillation does not grow in time and the error size associated with this is minor. 

Besides this numerical error, the accuracy of the mass and the area of the apparent horizon converge approximately at fourth order with respect to the grid spacing $\Delta x$. The numerical error for the mass and dimensionless spin increase approximately linearly in time, but for $\chi=0.8$ with $\Delta x\leq 0.016M_\mathrm{BH}$, which is the typical grid resolution of the present paper, the errors in mass and dimensionless spin are within $\approx 1.6$\% and $\Delta \chi \approx 0.004$, respectively, at $t=80,000M_\mathrm{BH}$. For $\Delta x=0.020M_\mathrm{BH}$, the error size is more than twice as large as that with $\Delta x =0.016M_\mathrm{BH}$. This illustrates that a sufficiently high grid resolution is necessary to accurately evolve the black hole. For model \texttt{AD20-7.8} with $\Delta x/M_\mathrm{BH,0} \approx 0.0215$, the grid resolution in the early stage of the black-hole evolution is so low that the mass and dimensionless spin are likely to be overestimated and underestimated, respectively. This is also the case for model \texttt{AD20x1}~\cite{Fujibayashi2022dec}. For this model the grid resolution for the early black-hole evolution was not so high that the black-hole mass and dimensionless spin were overestimated and underestimated, respectively. As a result, the specific angular momentum at the innermost stable circular orbit around the black hole was spuriously overestimated in the numerical computation, and thus, the matter around the black hole were more subject to falling into the black hole. This leaded to the overestimation of the black-hole mass and underestimation of the disk/torus mass. For this model, the NDAF phase was not found~\cite{Fujibayashi2022dec}, but this might be a spurious result due to the poor grid resolution. 

\section{Dependence on the grid resolution}\label{A3}

In this section, we compare the results of models \texttt{AD35-15} and \texttt{AD35-15-hi} as a convergence test. Figure~\ref{figapp2} shows the evolution of the mass and dimensionless spin (left) and the explosion energy and ejecta mass (right). We find a fair agreement between the results for different grid resolutions. For the black-hole mass, the higher-resolution results slightly in smaller mass. The primary reason for this is that with the higher-resolution, the viscous heating is more efficient, enhancing larger ejecta mass (see the right upper panel) while suppressing the accretion onto the black hole. Thus the black-hole mass presented in Fig.~\ref{fig:BH} may be slightly overestimated for their late stages while the ejecta mass may be underestimated in Fig.~\ref{fig:exp}.
The explosion energy are also slightly larger for the higher grid resolution, reflecting more energy injection from the viscous heating.  

\bibliography{reference}

\end{document}